\documentclass[a4paper,11pt]{article}
\pdfoutput=1
\usepackage{jheppub}
\usepackage{textcase}
\usepackage{graphicx,epsfig,psfrag,amssymb,hyperref}
\usepackage{multirow}
\usepackage{color,graphicx,epsfig,psfrag,amsmath,empheq}
\usepackage[dvipsnames]{xcolor}
\usepackage{bm}
\usepackage{mathrsfs,amsfonts,soul,color}
\usepackage[caption=false]{subfig}
\usepackage{hepunits}
\usepackage{enumitem}
\usepackage{placeins}
\usepackage{textcomp}
\usepackage{tcolorbox}
\usepackage{float}
\usepackage[toc,page]{appendix}
\usepackage[ruled,vlined]{algorithm2e}
\usepackage{wrapfig}

\definecolor{darkblue}{rgb}{0.1,0.1,1.0}

\begin{document}


\title{A Cautionary Tale of Decorrelating \\ Theory Uncertainties} 

\author[a,b]{Aishik Ghosh}
\author[b,c]{and Benjamin Nachman}

\affiliation[a]{Department of Physics and Astronomy, University of California, Irvine, CA 92697, USA}
\affiliation[b]{Physics Division, Lawrence Berkeley National Laboratory, Berkeley, CA 94720, USA}
\affiliation[c]{Berkeley Institute for Data Science, University of California, Berkeley, CA 94720, USA}

\emailAdd{aishikghosh@lbl.gov}
\emailAdd{bpnachman@lbl.gov}

\abstract{
A variety of techniques have been proposed to train machine learning classifiers that are independent of a given feature.  While this can be an essential technique for enabling background estimation, it may also be useful for reducing uncertainties.  We carefully examine theory uncertainties, which typically do not have a statistical origin.  We will provide explicit examples of two-point (fragmentation modeling) and continuous (higher-order corrections) uncertainties where decorrelating significantly reduces the apparent uncertainty while the actual uncertainty is much larger.  These results suggest that caution should be taken when using decorrelation for these types of uncertainties as long as we do not have a complete decomposition into statistically meaningful components.
}

\maketitle

\section{Introduction}
\label{sec:intro}

Modern machine learning classifiers hold great promise for increasing the sensitivity of high energy physics data analyses~\cite{Larkoski:2017jix,Guest:2018yhq,Albertsson:2018maf,Radovic:2018dip,Carleo:2019ptp,Bourilkov:2019yoi,Schwartz:2021ftp,Feickert:2021ajf}.  Typically, a classifier is trained using simulations and then the number of events passing a fixed threshold on the classifier in data and in simulation is counted.  A comparison between these counts is then used to estimate model parameters such as masses, couplings, and (new physics) cross sections (limits).  Theoretical and experimental uncertainties on the final result are accounted for by varying an aspect of the simulation and recomputing the predicted count using the nominal classifier.  The uncertainties in the model used for training affect the optimally of the classifier itself~\cite{Nachman:2019dol}, but typically do not cause a bias and can be accounted for~\cite{Ghosh:2021roe} by using parameterized classifiers~\cite{Cranmer:2015bka,Baldi:2016fzo}.

A variety of techniques have been proposed to render a classifier independent of a given feature~\cite{Louppe:2016ylz,Dolen:2016kst,Moult:2017okx,Stevens:2013dya,Shimmin:2017mfk,Bradshaw:2019ipy,ATL-PHYS-PUB-2018-014,DiscoFever,Xia:2018kgd,Englert:2018cfo,Wunsch:2019qbo,Rogozhnikov:2014zea,10.1088/2632-2153/ab9023,clavijo2020adversarial,Kasieczka:2020pil,Kitouni:2020xgb}.  This has become an essential tool for resonance searches, where thresholds on the classifier must not sculpt bumps in a given spectrum so that the Standard Model background can be estimated using sideband fits.  The same methodology has also been proposed to reduce systematic uncertainties.  If a classifier does not depend on a particular nuisance parameter, then the count computed when the parameter is varied will be the same as the nominal value.  This means that the uncertainty on the parameter(s) of interest will appear to be reduced. 

In the case that the systematic uncertainty is decomposed into its most fundamental components, each with a clear statistical interpretation, the above would be the end of the story.  The systematic uncertainty can be reduced through decorrelation and this would be useful if the classification performance does not rely strongly on the value of the nuisance parameters (otherwise, it may be better to profile instead~\cite{Ghosh:2021roe}).  However, theory uncertainties almost never satisfy these conditions.  These uncertainties are the result of approximations when performing calculations and are also due to parameter freedom in phenomenological models that are needed when first-principles calculations are not possible.  The canonical examples for these two types of uncertainties are perturbative uncertainties from series truncation and fragmentation modeling.  For the former, calculations are truncated at a fixed order in perturbation theory and the result depends on unphysical scales.  These scales are varied typically by factors of two in order to determine the uncertainty.  Fragmentation modeling uncertainties are often evaluated by comparing two different models, such as the string model~\cite{Andersson:1983ia,Sjostrand:1984ic} in the \textsc{Pythia}~\cite{Sjostrand:2006za,Sjostrand:2007gs} parton shower Monte Carlo (PSMC) and the cluster model~\cite{Webber:1983if,Winter:2003tt} in the \textsc{Herwig}~\cite{Bellm:2015jjp,Bahr:2008pv} PSMC.  These variations are then interpreted as a one standard deviation uncertainty and combined with other sources of uncertainty in a final statistical analysis.

\begin{figure}[h!]
\centering
\includegraphics[height=0.38\textwidth]{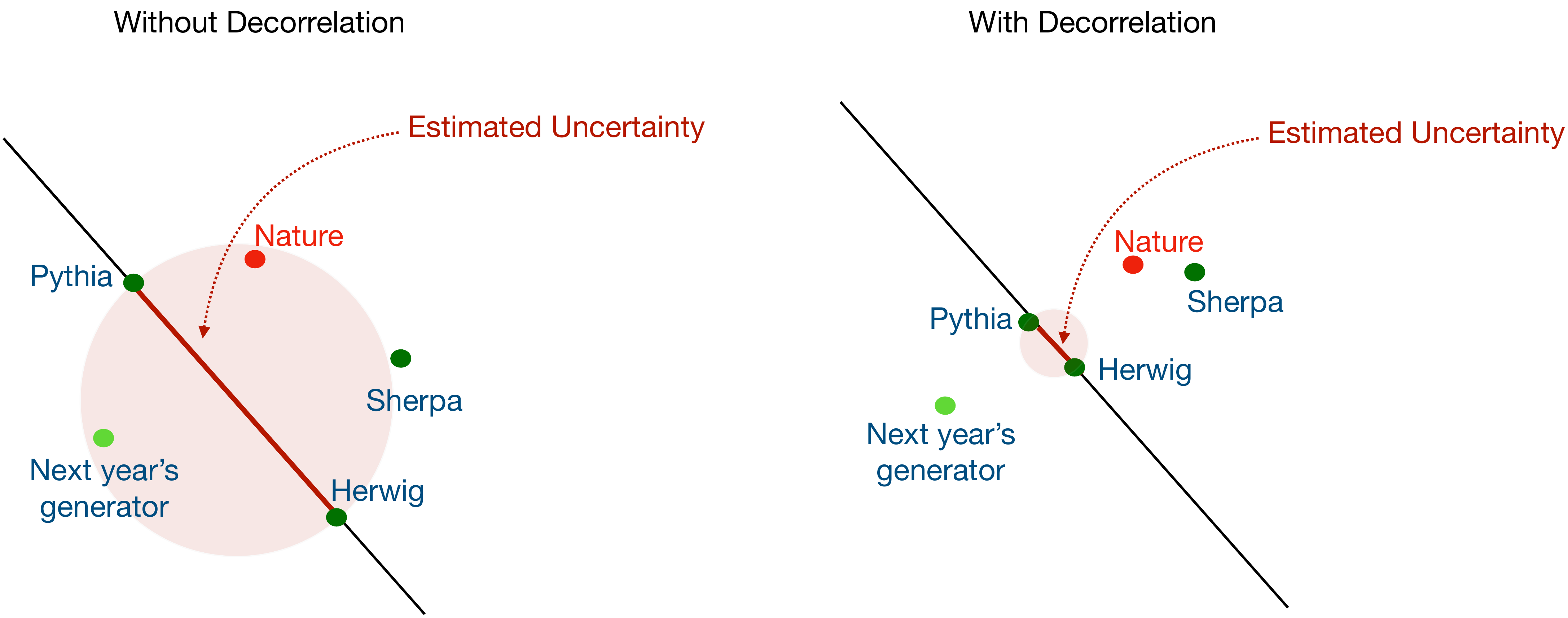}
\caption{An illustration of the potential impact of training a classifier to be decorrelated to two-point uncertainties. The distance between \textsc{Pythia} and \textsc{Herwig} is treated as the uncertainty. Left: Without decorrelation, the uncertainty covers nature even if nature does not lie on the line connecting \textsc{Pythia} and \textsc{Herwig}. Right: The distance between \textsc{Pythia} and \textsc{Herwig} is reduced due to the decorrelation requirement, resulting in a smaller estimate of the uncertainty, which no longer covers nature. These diagrams are meant only to be intuitive illustrations.}
\label{fig:intuitiveIllustration}
\end{figure}

We examine the interplay of decorrelation with theory uncertainties.  In particular, we will show that constructing a classifier that is independent of a given theory nuisance parameter does not mean that the theory uncertainty is zero.  Instead, it means that the only handle to determine the theory uncertainty is eliminated.  Figure~\ref{fig:intuitiveIllustration} illustrates the intuition behind why this might be the case. As concrete examples, we study fragmentation modeling for Lorentz-boosted $W$ boson jet classification and factorization scale variations when classifying events as either from $W$+jets or $t$-channel single top quark events.

This paper is organized as follows.  Section~\ref{sec:dec} briefly introduces existing decorrelation techniques.  Numerical examples of both two-point and continuous uncertainties are provided in Sec.~\ref{sec:examples}.  The paper ends with conclusions and outlook in Sec.~\ref{sec:conclusion}.

\section{Decorrelation Techniques}
\label{sec:dec}

Let $x\in\mathbb{R}^n$ be the features used for classification.  Suppose that there is a feature\footnote{This also applies to cases where $m$ is multi-dimensional, but we restrict to the one-dimensional setting here for simplicity and because it is widely used.} $m\in\mathbb{R}$ that we want to be decorrelated from a classifier $f(x):\mathbb{R}^n\rightarrow\mathbb{R}$.  One can achieve this decorrelation by minimizing the following loss functional $L$:

\begin{align}
\label{eq:dec}
    L[f(x)]=\sum_{i\in S}L_\text{class}(f(x_i),1)+\sum_{i\in B} w(m_i) L_\text{class}(f(x_i),0)+\lambda\,\sum_{i\in B}L_\text{decor}(f(x_i),m_i)\,,
\end{align}
where $S$ and $B$ represent signal and background events, respectively.  The loss $L_\text{class}$ is the classifier loss and is often the binary cross entropy $L_\text{class}(f(x),y) = y\log(f(x))+(1-y)\log(1-f(x))$.  The function $w(m)$ represents a weighting function and $\lambda$ represents a hyperparameter that controls the strength of the decorrelation.  Finally, $L_\text{decor}$ is a term that penalizes any dependence between $f$ and $m$.  This last term in Eq.~\ref{eq:dec} is schematic as the decorrelation penalty often acts at the level of batches of events and not individual examples.  Standard classification corresponds to $w(m)=1$ and $\lambda=0$.  Decorrelation approaches include:

\begin{itemize} 
    \item Planing~\cite{Chang:2017kvc,1511.05190}: $\lambda=0$ and $w(m_i)\approx p_S(m)/p_B(m)$ so that the marginal distribution of $m$ is non-discriminatory after the reweighting.
    \item Adversaries~\cite{Louppe:2016ylz,Shimmin:2017mfk,Englert:2018cfo,clavijo2020adversarial}:  $w(m)=1$, $\lambda<0$, and $L_\text{decor}$ is the loss of a second neural network (adversary) that takes $f(x)$ as input and tries to learn some properties of $m$.
    \item Distance Correlation (DisCo)~\cite{DiscoFever,Kasieczka:2020pil}: $w(m)=1, \lambda>0$, and the last term in Eq.~\eqref{eq:dec} is the \textit{distance correlation}~\cite{szekely2007, szekely2009, SzeKely:2013:DCT:2486206.2486394,szekely2014} between $f(x)$ and $m$ for the background.
    \item Flatness~\cite{Rogozhnikov:2014zea}: $w(m)=1$, $\lambda>0$, and $L_\text{decor}=\sum_m b_m \int |F_m(s)-F(s)|^2\, {\rm d}s$ where the sum runs over mass bins, $b_m$ is the fraction of candidates in bin $m$, $F$ is the cumulative distribution function, and $s=f(x)$ is the classifier output.  This is generalized to Moment Decorrelation (MoDE) in Ref.~\cite{Kitouni:2020xgb} to allow for a given dependence of $f$ on $m$.
\end{itemize}

In the examples below, we focus on the adversarial case as it is the most explored in the literature.  However, the same ideas apply to all decorrelation methods.

\section{Numerical Examples}
\label{sec:examples}

All neural networks are implemented using \textsc{Keras}~\cite{keras} with the \textsc{Tensorflow} backend~\cite{tensorflow} and optimized with \textsc{Adam}~\cite{adam}.

\subsection{Two-point Uncertainty: Fragmentation Modeling}
\label{sec:twopoint}

General purpose event generators use perturbation theory when they can and phenomological models to describe non-perturbative effects such as hadronization.  The standard procedure for estimating the uncertainty due to the model choice is to compare the predictions from two different models.  This uncertainty is typical largest when the analysis strategy exploits subtle correlations in the high-dimensional radiation pattern.  For example, tagging the origin of high $p_T$ jets is a widely-studied scenario~\cite{ATLAS:2018wis,CMS:2020poo,Kasieczka:2019dbj} for machine learning whereby the detailed jet substructure can be used for classification.  In this section, we study Lorenz-boosted $W$ boson tagging, where the signal is hadronically decaying, high $p_T$ $W$ bosons and the background is generic quark and gluon jets.  A single large-radius jet is often sufficient to capture most of the $W$ boson decay products and its two-prong substructure is distinct from typical quark and gluon jets.

Samples were generated with MadGraph5\_aMC@NLO 2.7.3~\citep{Alwall:2014hca} for modeling $pp$ collisions at $\sqrt{s}$ = 13 TeV. The {\texttt{NNPDF23\_nlo\_as\_0118}} \citep{Ball:2012cx} parton distribution function is used. The hard-scattering events are passed to \textsc{Pythia 8.303}~\citep{Sjostrand:2007gs} to simulate the parton shower and hadronization, using the default settings. \textsc{Herwig} 7.2.2~\citep{Bellm:2015jjp} with angularly-ordered showers and \textsc{Sherpa} 2.2.2~\citep{Gleisberg:2008ta,Sherpa:2019gpd} with default settings are also used to model the parton shower and hadronization\footnote{While \textsc{Herwig} and \textsc{Sherpa} both use a cluster model for fragmentation, the actual \textsc{Sherpa} implementation is based on \cite{Winter:2003tt} and differs from \textsc{Herwig} in several respects.}. The jets are clustered by \textsc{Pyjet}~\citep{noel_dawe_2021_4446849,Cacciari:2011ma} and the anti-$k_t$~\citep{Cacciari:2008gp} algorithm with radius parameter $R$ = 1.2.

\begin{figure}[h!]
\centering
\includegraphics[height=0.28\textwidth]{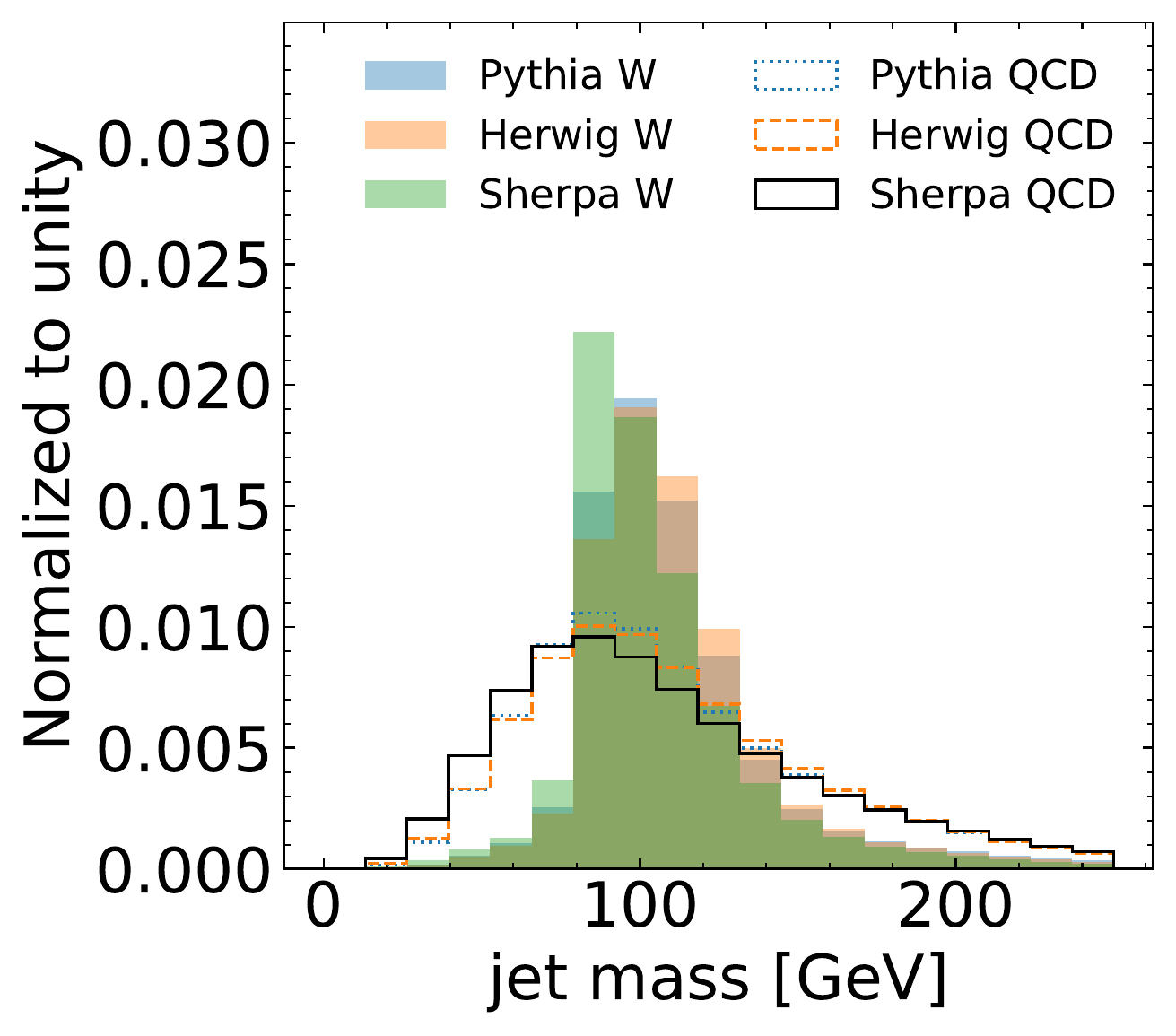}\includegraphics[height=0.28\textwidth]{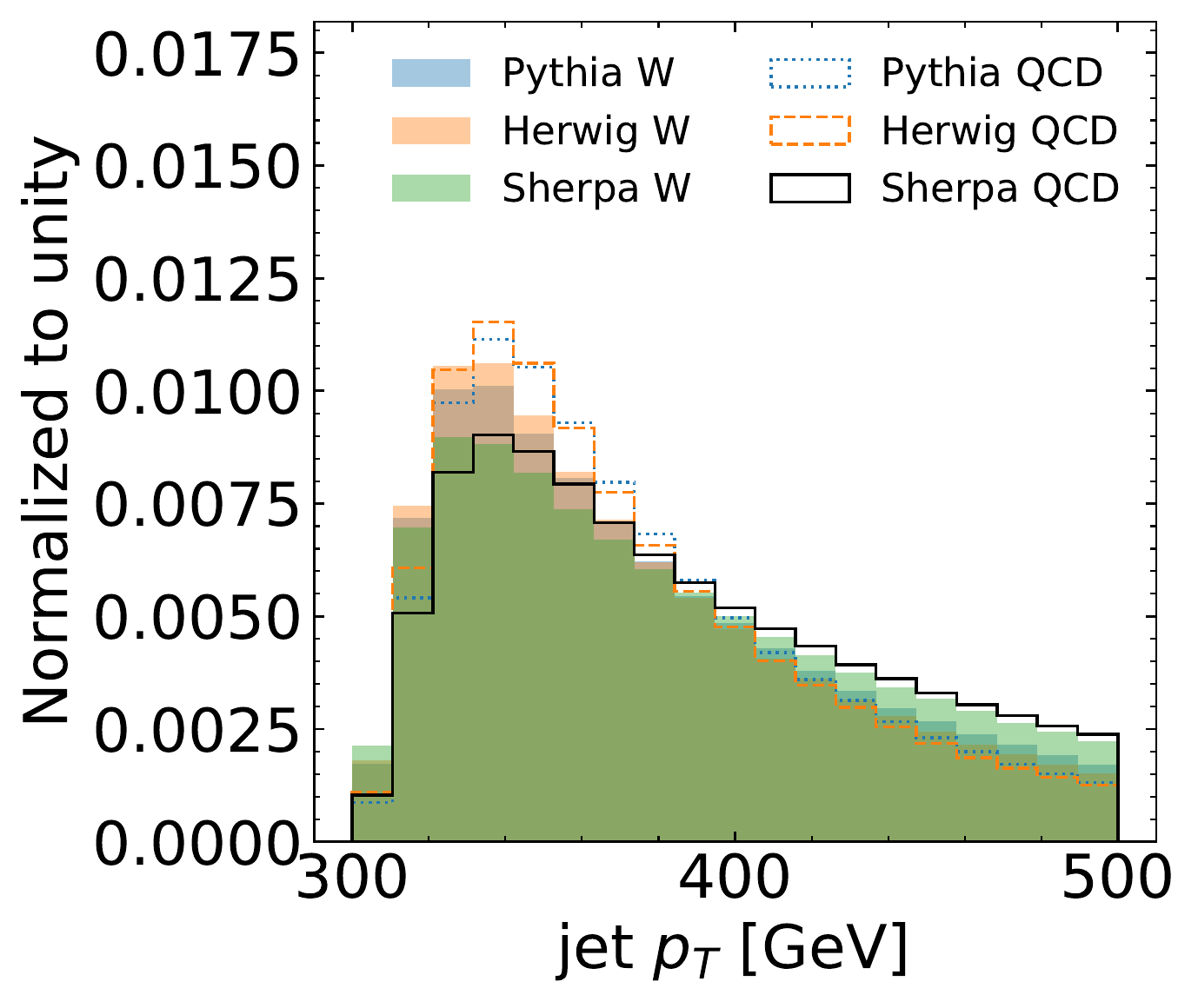}\includegraphics[height=0.28\textwidth]{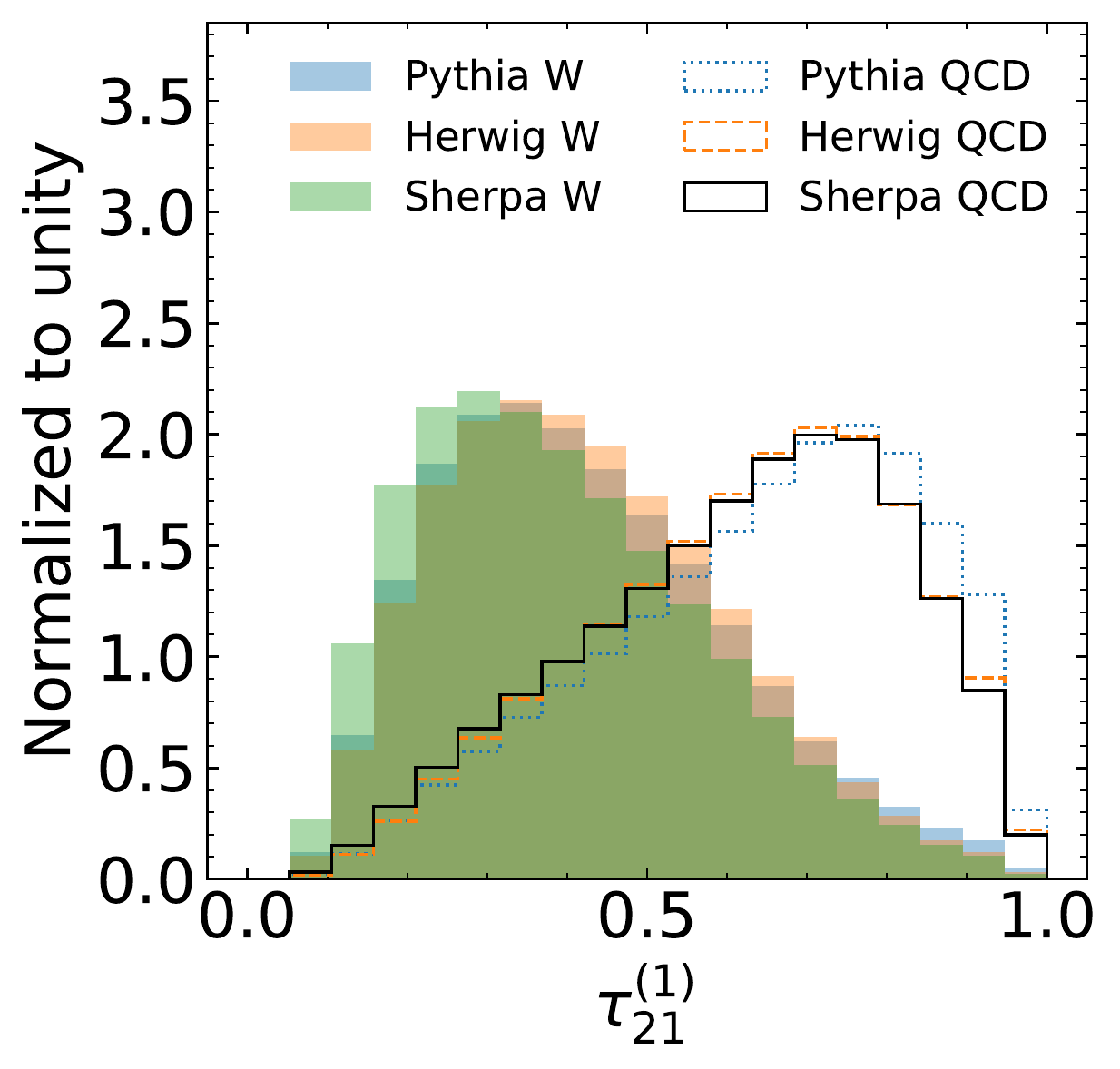}\\
\includegraphics[height=0.28\textwidth]{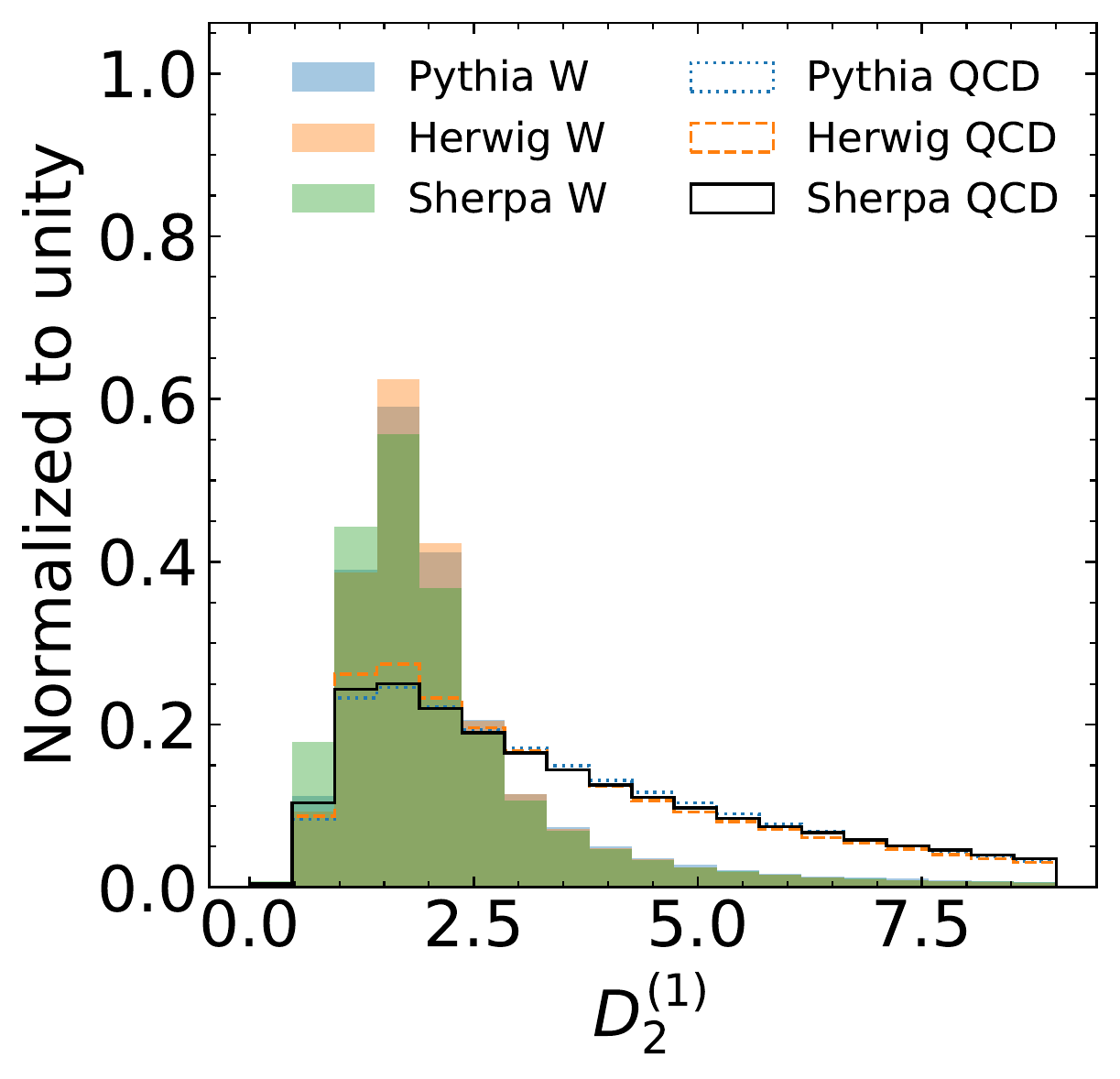}\includegraphics[height=0.28\textwidth]{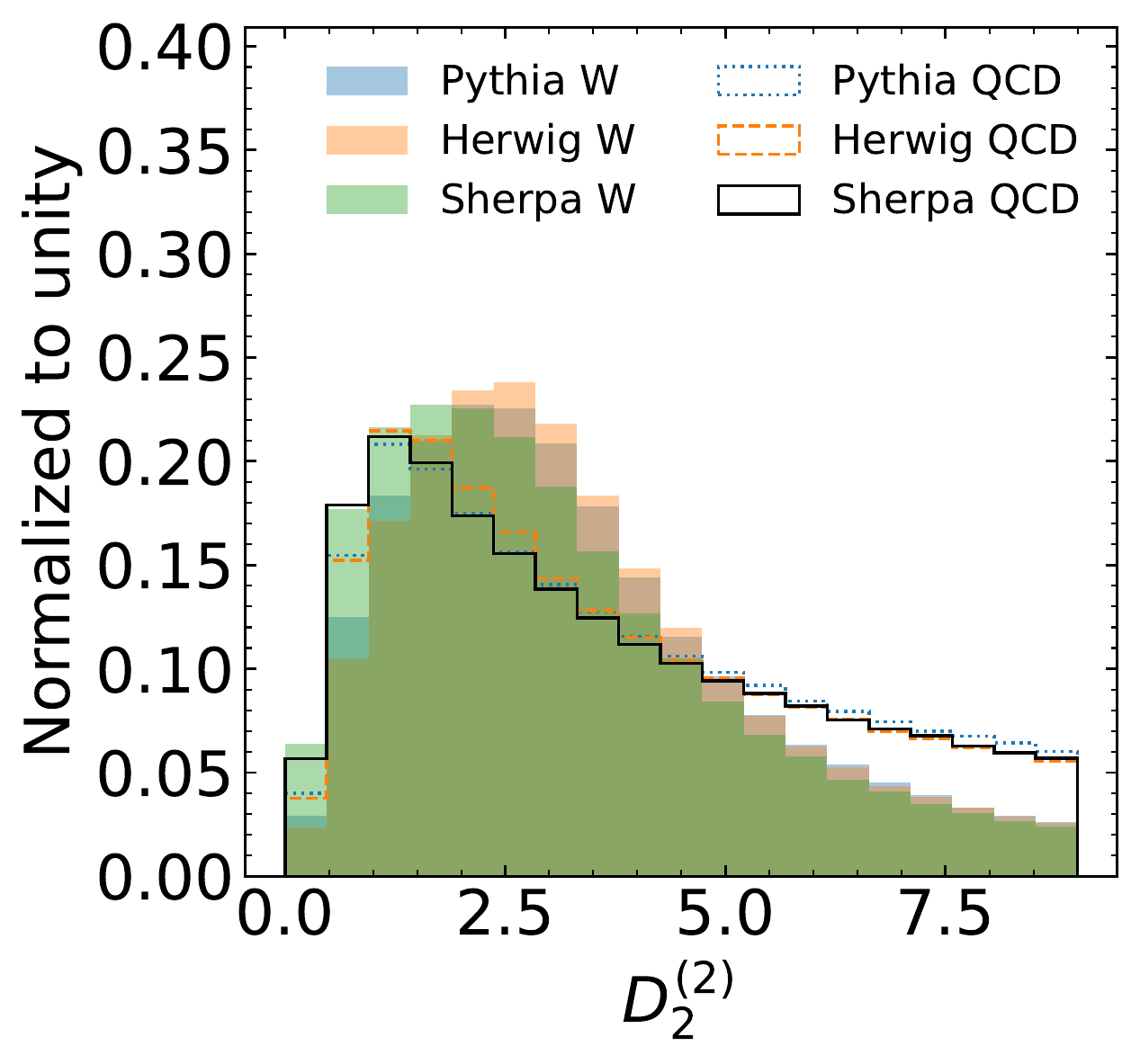}\includegraphics[height=0.28\textwidth]{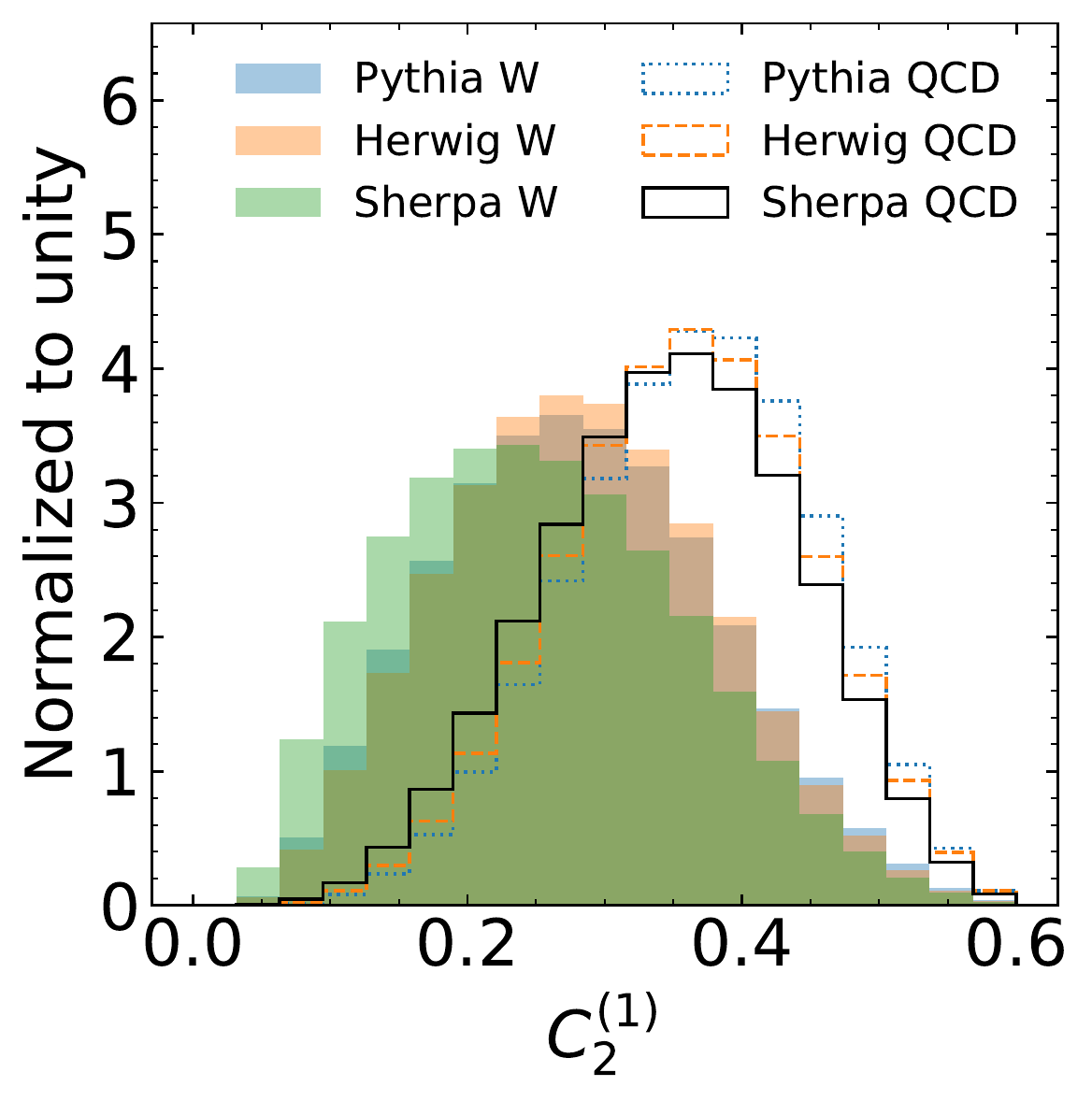}\\
\includegraphics[height=0.28\textwidth]{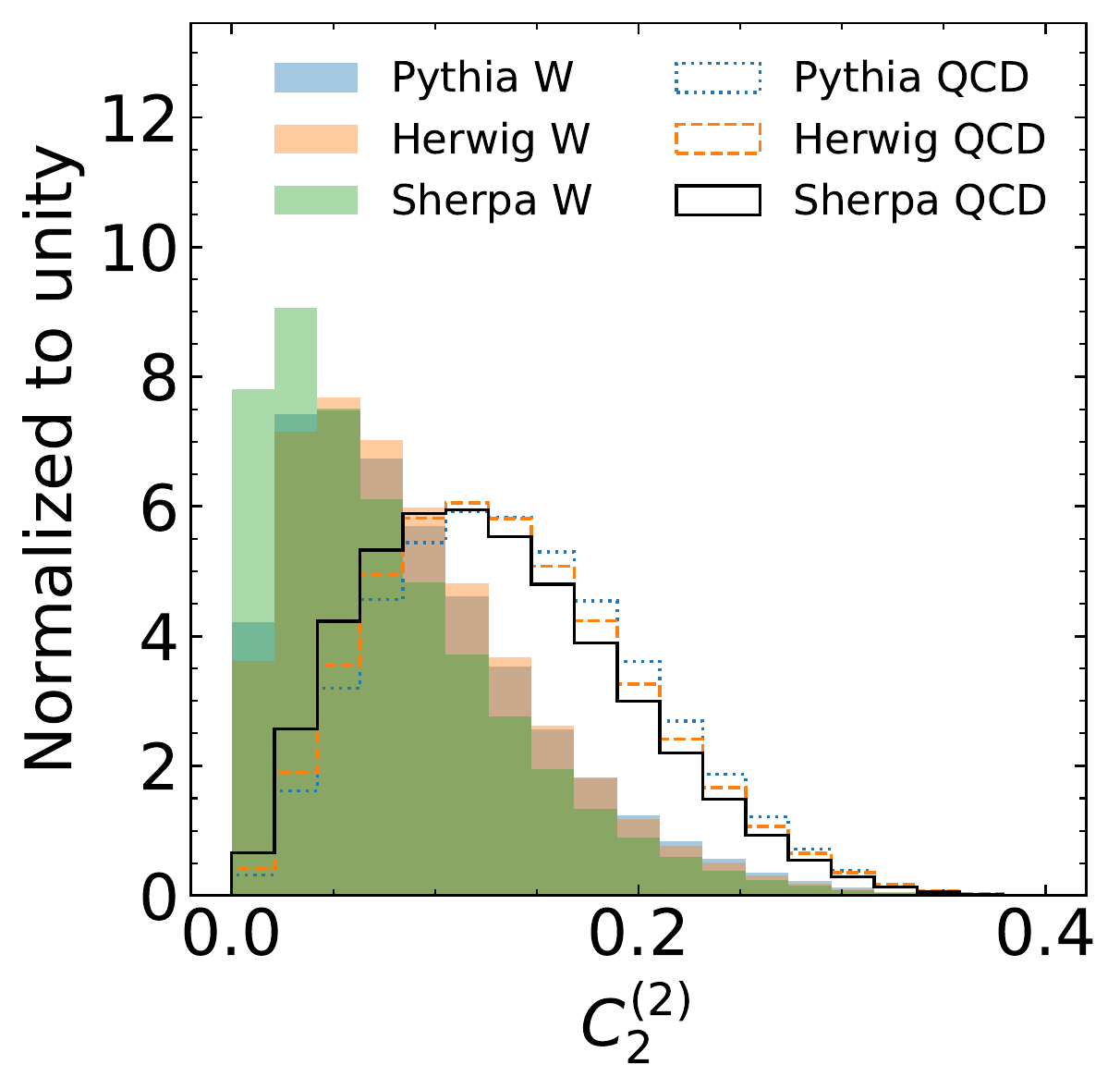}
\caption{The seven inputs used to train a classifier to distinguish boosted $W$ boson jets from generic QCD jets events.}
\label{fig:winputs}
\end{figure}

A set of high-level jet substructure features are used to distinguish $W$ jets from QCD jets.  These features are illustrated in Fig.~\ref{fig:winputs} and briefly described in the following.  The kinematics are probed with the jet mass and transverse momentum.  Jet substructure observables include $n$-subjettiness ratio $\tau_{21}=\tau_2/\tau_1$~\cite{Thaler:2010tr,Thaler:2011gf}, and energy correlation function ratios $D_2^{(\beta)}=e_3^{(\beta)}/(e_2^{(\beta)})^3$~\cite{Larkoski:2014gra} and $C_2^{(\beta)}=e_3^{(\beta)}/(e_2^{(\beta)})^2$~\cite{Larkoski:2013eya}, where $e_i$ is the normalized sum over doublets ($i=2$) or triplets ($i=3$) of constituents inside jets, weighted by the product of the constituent transverse momenta and pairwise angular distances.  For this analysis, we consider both $\beta=1$ and $\beta=2$.  

As expected, the mass peaks near the $W$ boson mass of 80 GeV~\cite{10.1093/ptep/ptaa104} for the signal and has a broad distribution for the background.  The signal peak is slightly higher than the $W$ boson mass due to underlying event and other event contamination.  This could be mitigated with grooming~\cite{Butterworth:2008iy,Ellis:2009me,Krohn:2009th,Dasgupta:2013ihk,Larkoski:2014wba}.  The jet $p_T$ is not very discriminating by construction.  The two-prong nature of the signal jets is quantified by a low $\tau_{21}, D_2$, and $C_2$.

\begin{figure}[h!]
\centering
\includegraphics[height=0.65\textwidth]{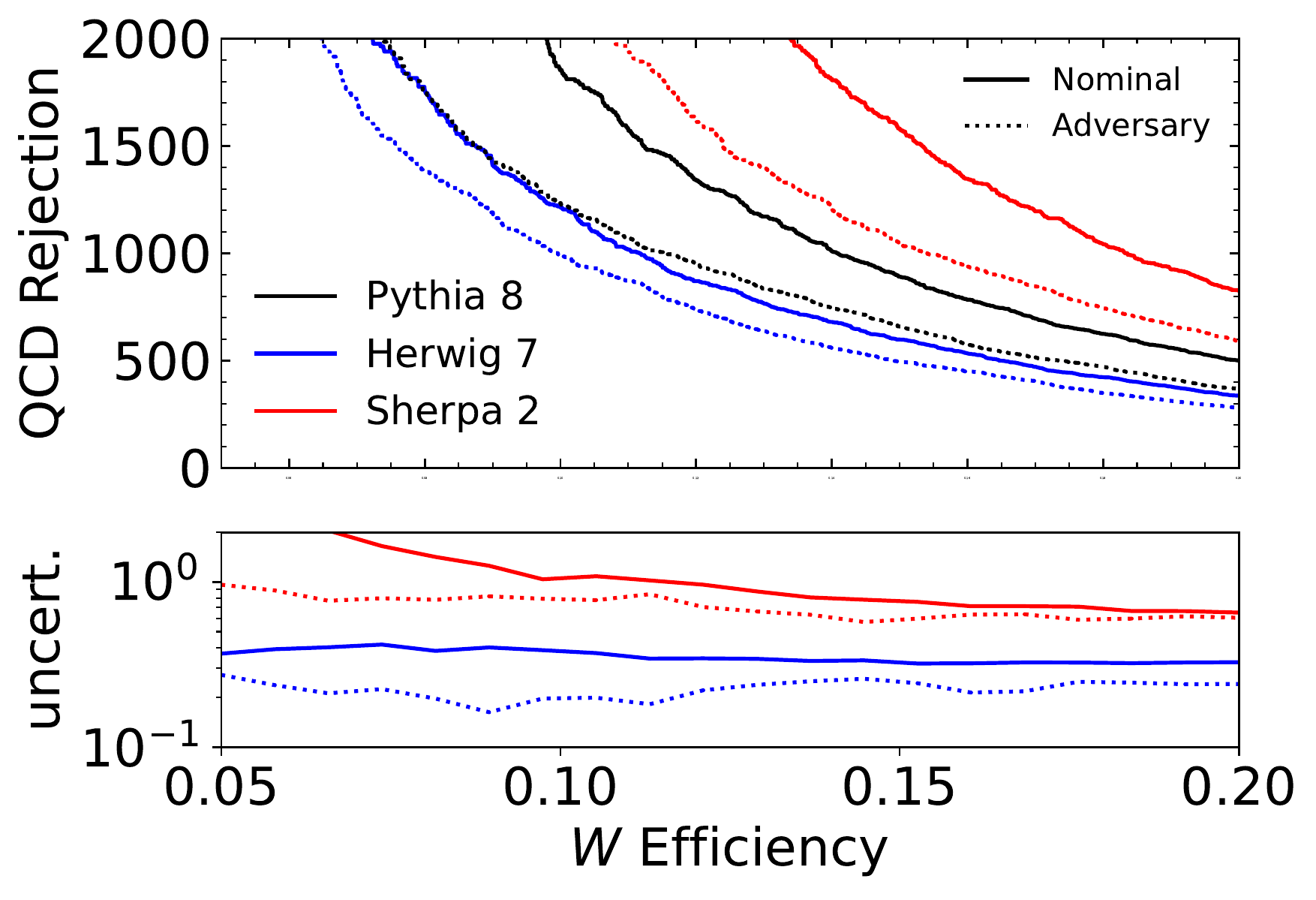}
\caption{The QCD rejection (inverse QCD efficiency) as a function of the $W$ jet efficiency for classifiers applied to \textsc{Pythia}, \textsc{Herwig}, and \textsc{Sherpa} jets.  The solid lines correspond to the nominal classifier trained with \textsc{Pythia} while the dotted lines correspond to the adversarial setup that uses both \textsc{Pythia} and \textsc{Herwig}.  The bottom panel shows the relative absolute difference with respect to \textsc{Pythia} (nominal or adversarial, as appropriate). Note that the lower panel has a logarithmic vertical axis.  While adversarial training reduces the difference in performance between \textsc{Pythia} and \textsc{Herwig}, the difference to \textsc{Sherpa} remains large, indicating that the the true uncertainty will be underestimated if a third independent sample is unavailable.}
\label{fig:rocs}
\end{figure}

A classifier is trained using the seven features presented in Fig.~\ref{fig:winputs} to distinguish $W$ jets from QCD jets.  The nominal classifier is trained using the \textsc{Pythia} simulation and is parameterized as a neural network with two hidden layers of 50 nodes each.  Rectified Linear Unit (ReLU) activations are used for the intermediate layers and the final output is passed through a sigmoid function.  The binary cross-entropy is used for training with a batch size of 100 and for 20 epochs.  None of these parameters were optimized, although minor variations were found to have little impact on performance.  The performance of this nominal classifier evaluated on \textsc{Pythia}, \textsc{Herwig}, and \textsc{Sherpa} is shown in Fig.~\ref{fig:rocs}.  We focus on the region near 10-15\% signal efficiency, which is a typical working point for LHC analyses. In this range, the background rejection (inverse QCD efficiency) is between a few hundred and a few thousand.  

A second network is trained as part of an adversarial approach.  This second network uses both \textsc{Pythia} and \textsc{Herwig} events and minimizes the following loss:

\begin{align}
\label{eq:advloss}\nonumber
    L[f,g]=&-\left(\sum_{i\in W} \log(f(x_i))-\sum_{i\in\text{QCD}}\log(1-f(x_i))\right)\\
    &+ \lambda \left(\sum_{i\in \text{Pythia}} \log(g(f(x_i),y_i))-\sum_{i\in\text{Herwig}}\log(1-g(f(x_i),y_i))\right)\,,
\end{align}
where $y_i=0$ for $W$ jets and $y_i=1$ for QCD jets.  Furthermore, $\lambda=10$.  Note that unlike Eq.~\ref{eq:dec}, Eq.~\ref{eq:advloss} has the labels as part of the function for the adversary.  This means that the labels for the classifier are given as an input feature to the adversary, which allows the adversary to potentially learn separate decision functions for $W$ jets and QCD jets.  
The classifier network $f$ has the same composition as the nominal classifier described above: two hidden layers with 50 nodes each.  The adversary has five hidden layers with 50 nodes each.  As $W$ jets are more different from QCD jets than \textsc{Pythia} jets are from \textsc{Herwig} jets, the adversary has a more difficult task, which is why $g$ has a more complex architecture.  It was found that adding the label $y_i$ to $g$ as well as multiplying the gradient for the adversary by 10 improved performance and stability.
The minimax nature of the optimization in Eq.~\ref{eq:advloss} is implemented by connecting the adversary to the classifier via a gradient reversal layer~\cite{pmlr-v37-ganin15} that multiplies the gradient by a fixed negative constant during backpropagation.
The classifier network is then extracted after training for 20 epochs.  When $\lambda=0$, the performance was found to be the same as for the nominal case\footnote{Note that when $\lambda=0$, the adversarial setup is slightly different than the nominal configuration because both \textsc{Pythia} and \textsc{Herwig} are used for training.  This has little impact on the results - see Appendix~\ref{sec:lamb0}.}.

Figure~\ref{fig:rocs} shows that the performance of the adversarially trained classifier is worse than the nominal case.  This drop in performance is the cost for building a classifier that is insensitive to fragmentation model variations.  The difference between \textsc{Pythia} and \textsc{Herwig} for the nominal classifier is about 40\% at 10\% $W$ efficiency while it is only about 20\% for the adversarially trained network\footnote{It is possible this could be reduced with further hyperparameter tuning.  We found some parameters that made this smaller, but with significant variation across trainings.  The configuration reported here was found to be robust to retraining.}. The reduced difference may give the impression that the adversarially trained classifier has successfully learnt to be less sensitive to fragmentation model variations. However, the difference between \textsc{Sherpa} and \textsc{Pythia} is nearly the same for the nominal and the adversarially trained classifier.  This means that the `true' uncertainty would be significantly underestimated if only \textsc{Pythia} and \textsc{Herwig} were available.  It is often the case that only two fragmentation models are available.  

\subsection{Continuous Uncertainty: Higher-order Corrections}
\label{sec:cont}

The uncertainty from truncating the order of a perturbative calculation is typically estimated by varying the unphysical scales. Usually, there are renormalization scale and factorization scale uncertainties.  For simplicity, we focus here on the factorization scale, which dictates the separation between long- and short-distance physics.  The standard procedure is to set the factorization scale to the typical momentum transfer in the problem.

To study the impact of factorization scale variations, we consider measurements of $t$-channel single top quark production.  One of the main backgrounds for this process is $W$+jets production and machine learning is already used by ATLAS~\cite{ATLAS:2016qhd} and CMS~\cite{CMS:2019jjp} to enhance the signal.  The semileptonic channel is studied as it has a much smaller background than the all-hadronic channel.  The final state is characterized by an isolated lepton, missing transverse momentum, and jets.

Events are simulated using MadGraph5\_aMC@NLO (MG5\_aMC)~3.1.1~\cite{Alwall:2014hca} interfaced with \textsc{Pythia} 8.244~\cite{Sjostrand:2007gs} for the parton shower and \textsc{Delphes}~3.4.2~\cite{deFavereau:2013fsa,Mertens:2015kba,Selvaggi:2014mya} for detector simulations with the default CMS card.  Particle flow candidates are used as inputs to jet clustering, implemented using \textsc{FastJet}~3.2.1~\cite{Cacciari:2011ma,Cacciari:2005hq} and the anti-$k_t$ algorithm~\cite{Cacciari:2008gp} with radius parameter $R=0.5$.  For simplicity, $W$ bosons are forced to decay into muons and events are required to have at least one isolated and identified muon using the default reconstruction algorithm in \textsc{Delphes}.  Usually, one uses the highest precision method possible and then scale variations give the uncertainty from the finite truncation of the perturbative series.  In order to compare with the `true' uncertainty, we artificially truncate the series early and then use the higher-order calculation as the reference uncertainty.  In particular, the nominal simulation is performed at leading order (LO) in the strong coupling constant and then an additional sample for the $t$-channel process is simulated at next-to-leading order (NLO).

\begin{figure}[h!]
\centering
\includegraphics[height=0.25\textwidth]{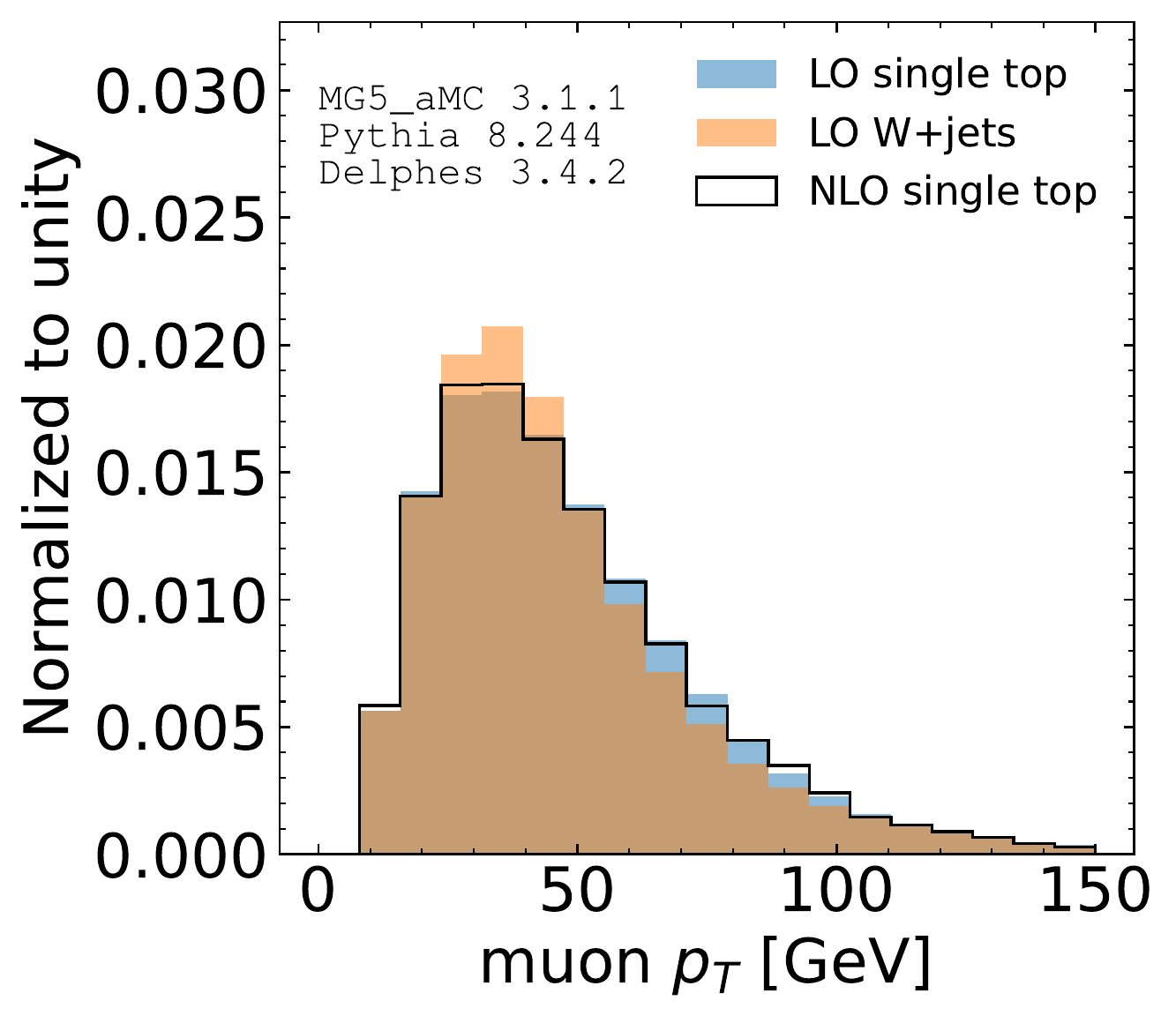}\includegraphics[height=0.25\textwidth]{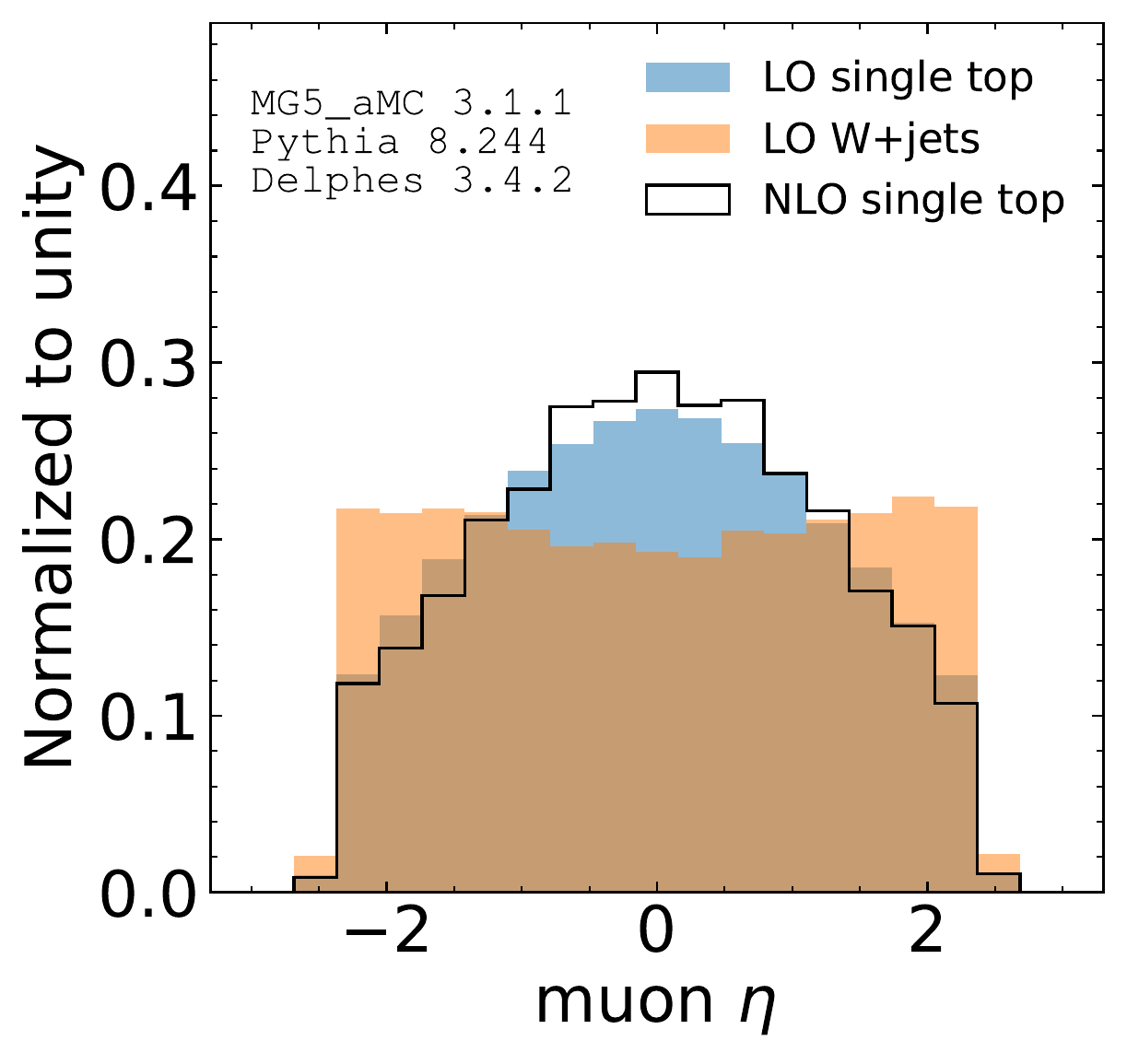}\includegraphics[height=0.25\textwidth]{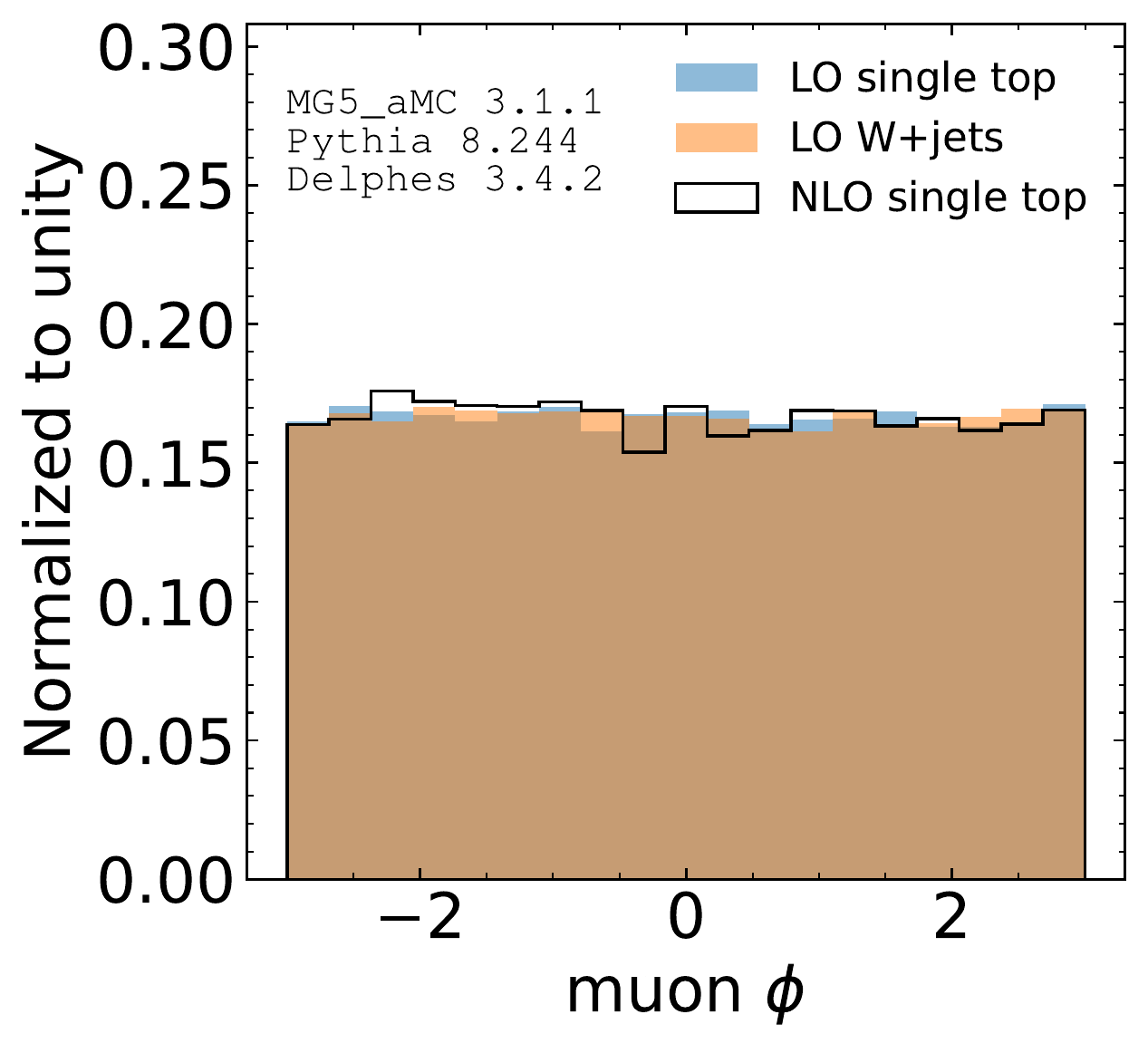}\\
\includegraphics[height=0.25\textwidth]{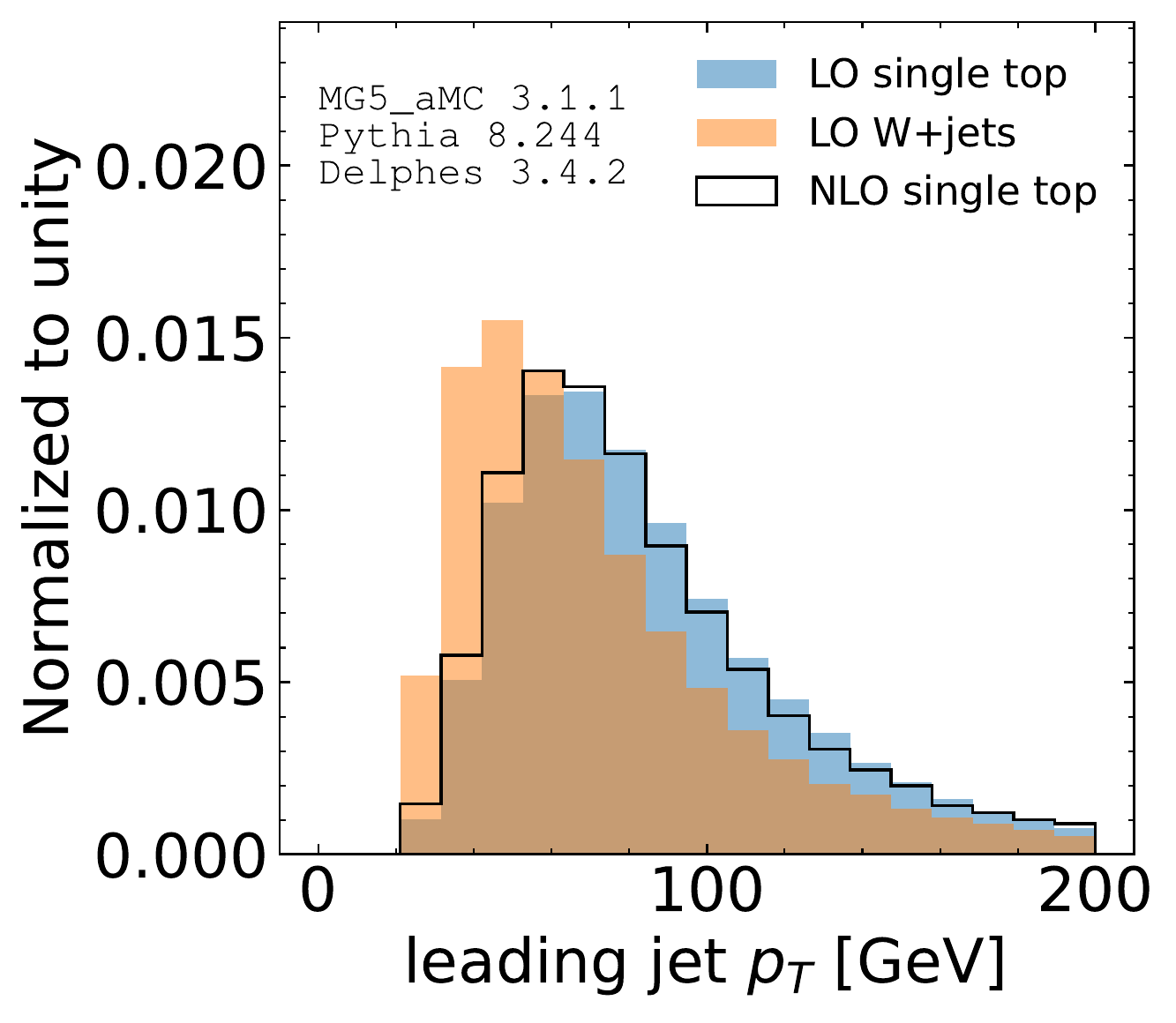}\includegraphics[height=0.25\textwidth]{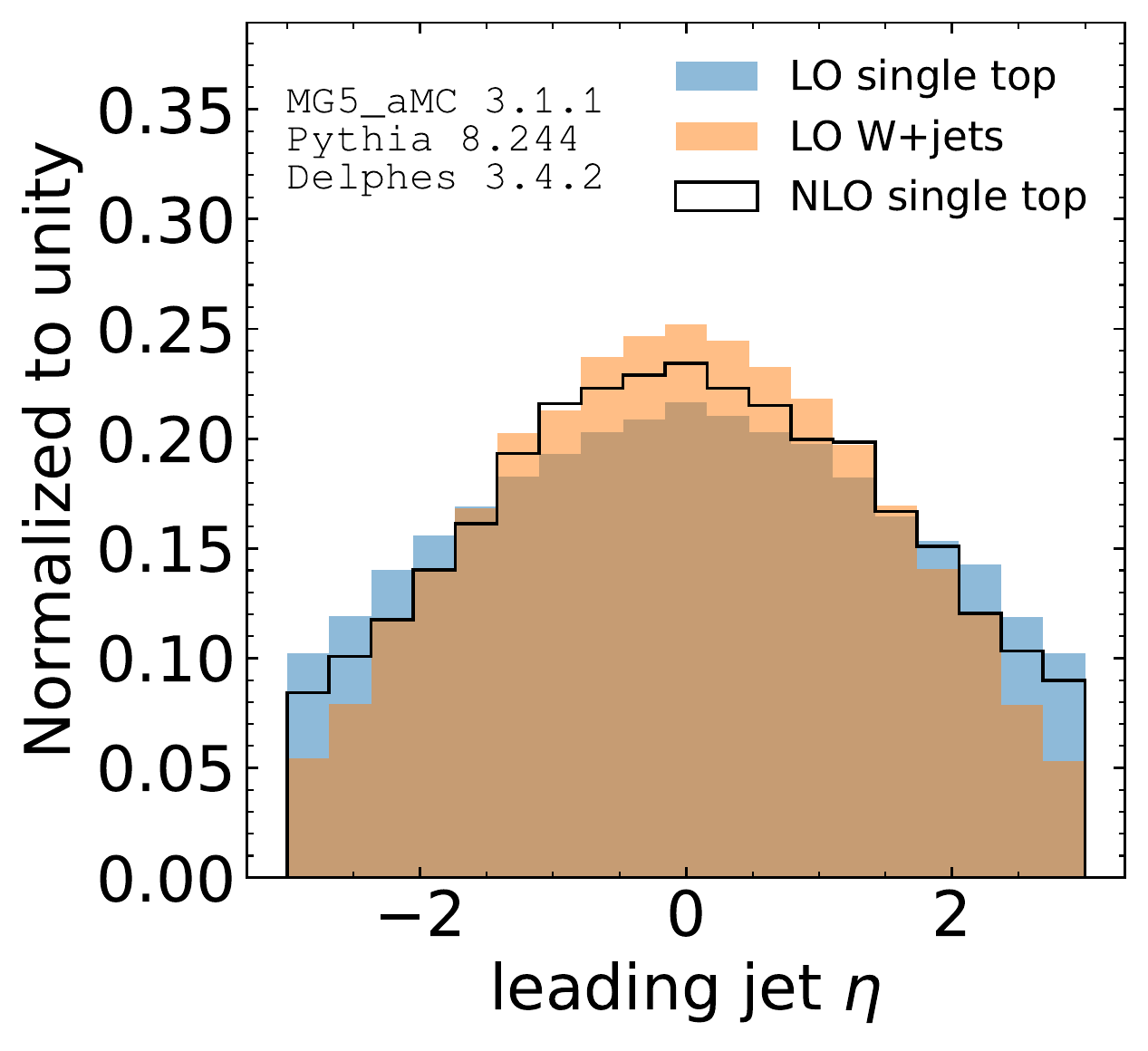}\includegraphics[height=0.25\textwidth]{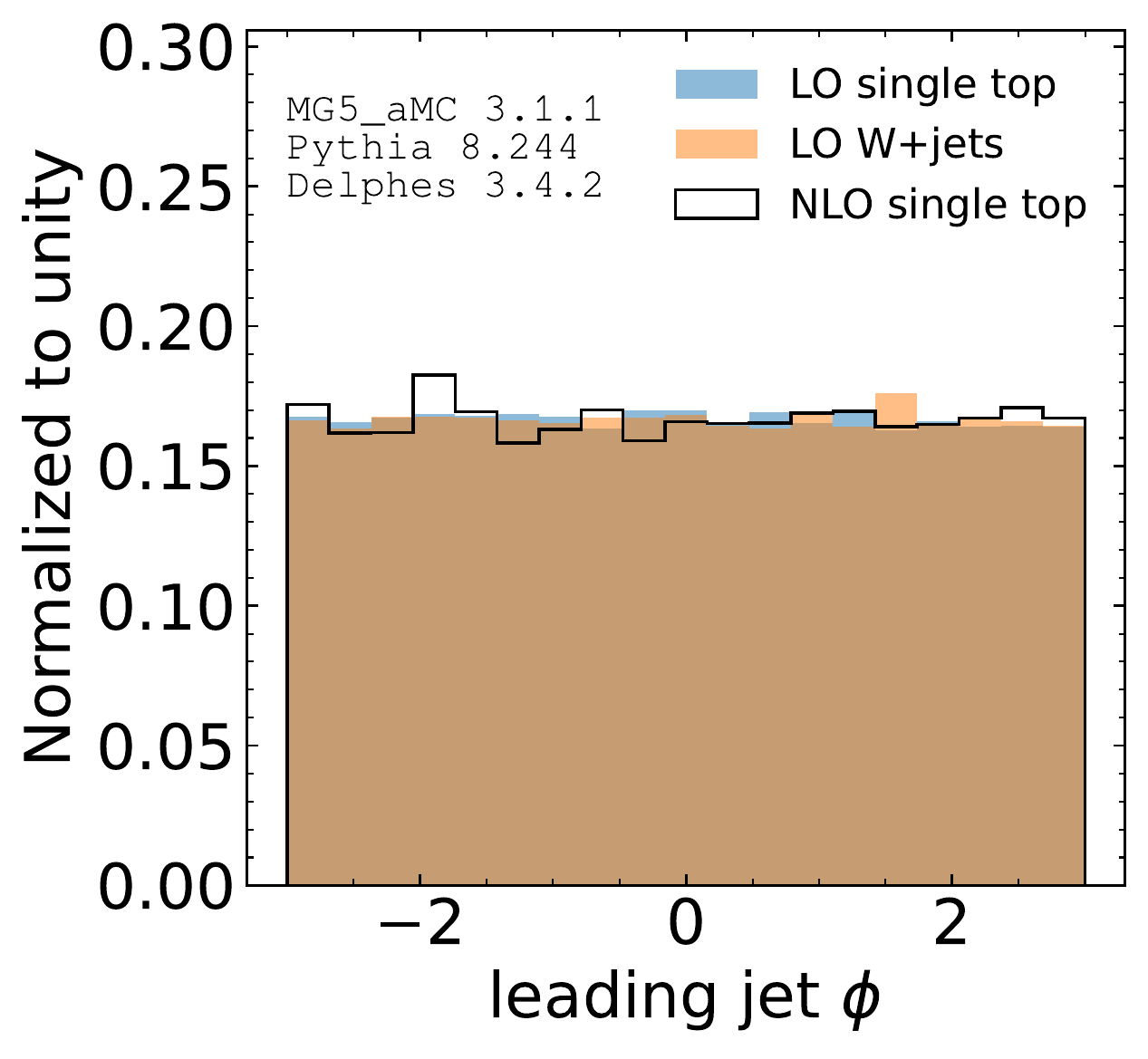}\\
\includegraphics[height=0.25\textwidth]{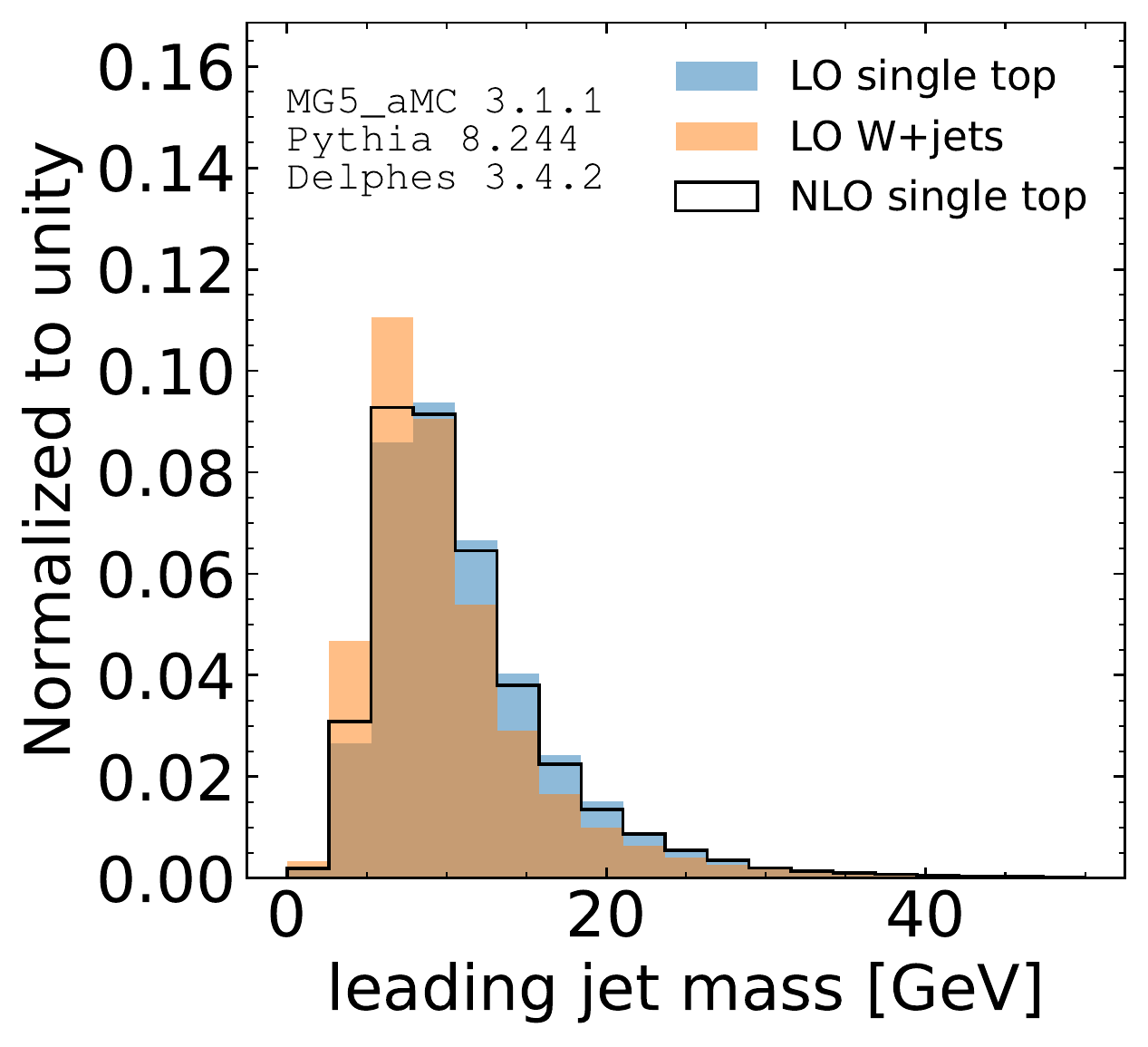}\includegraphics[height=0.25\textwidth]{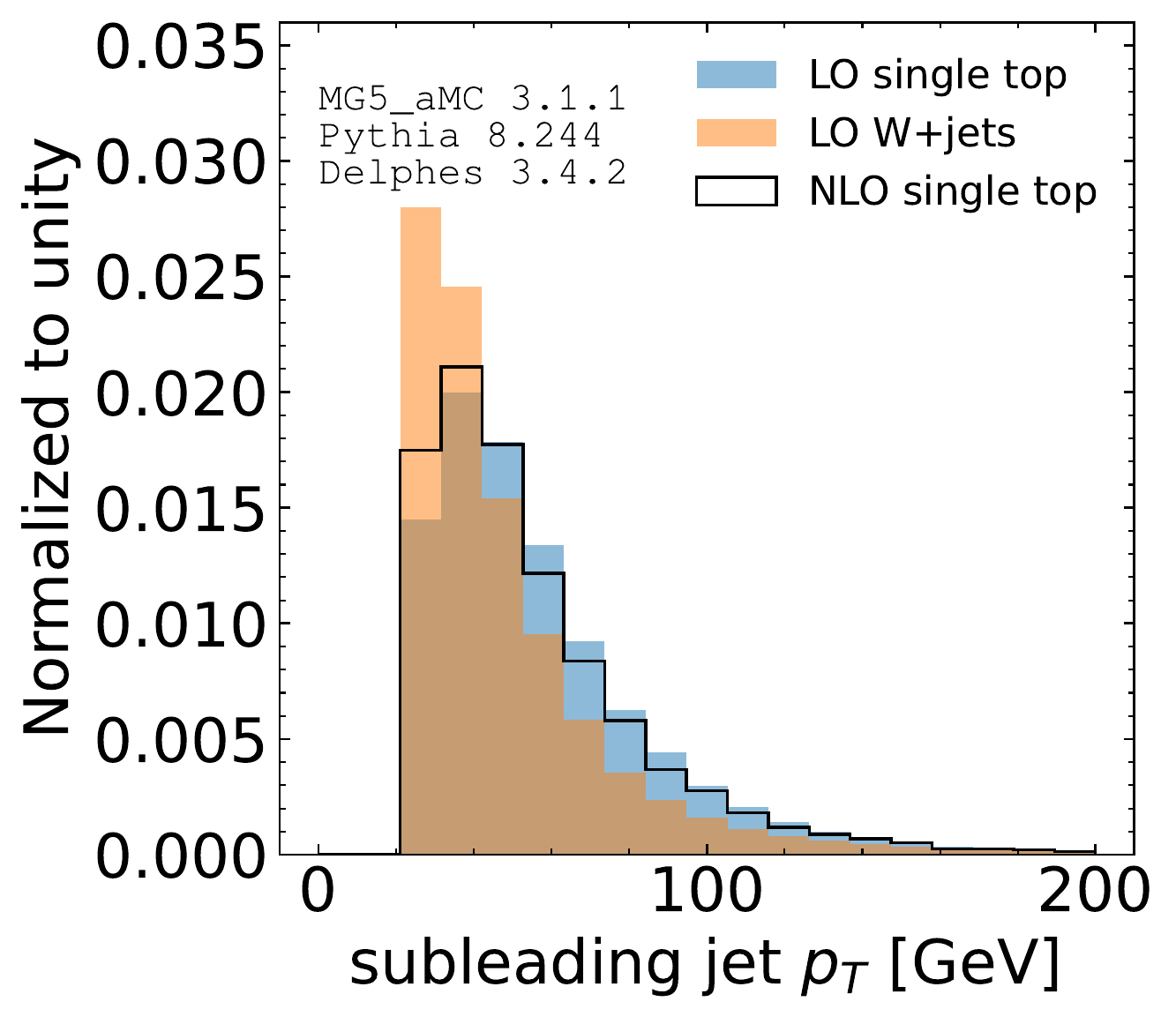}\includegraphics[height=0.25\textwidth]{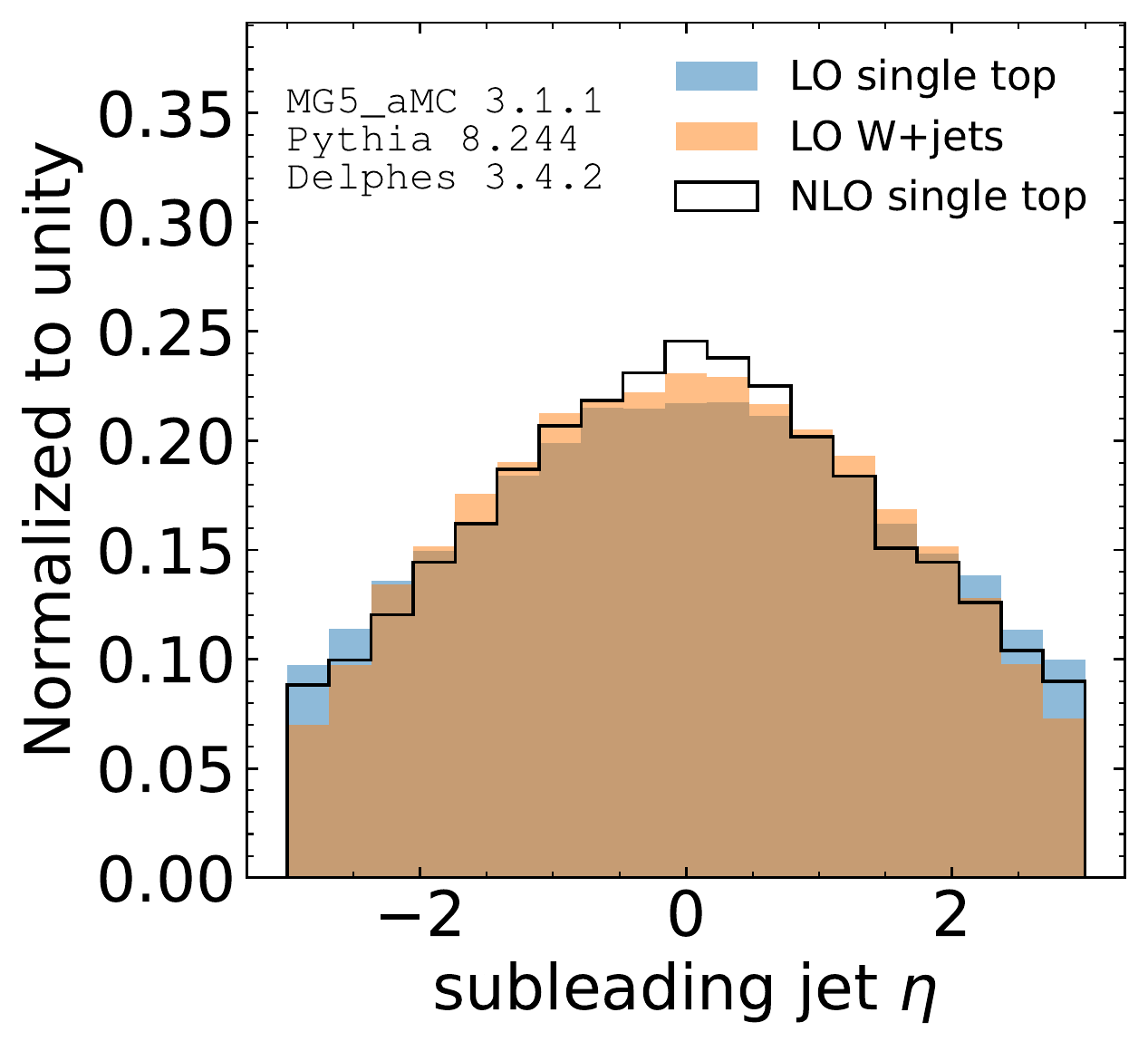}
\includegraphics[height=0.25\textwidth]{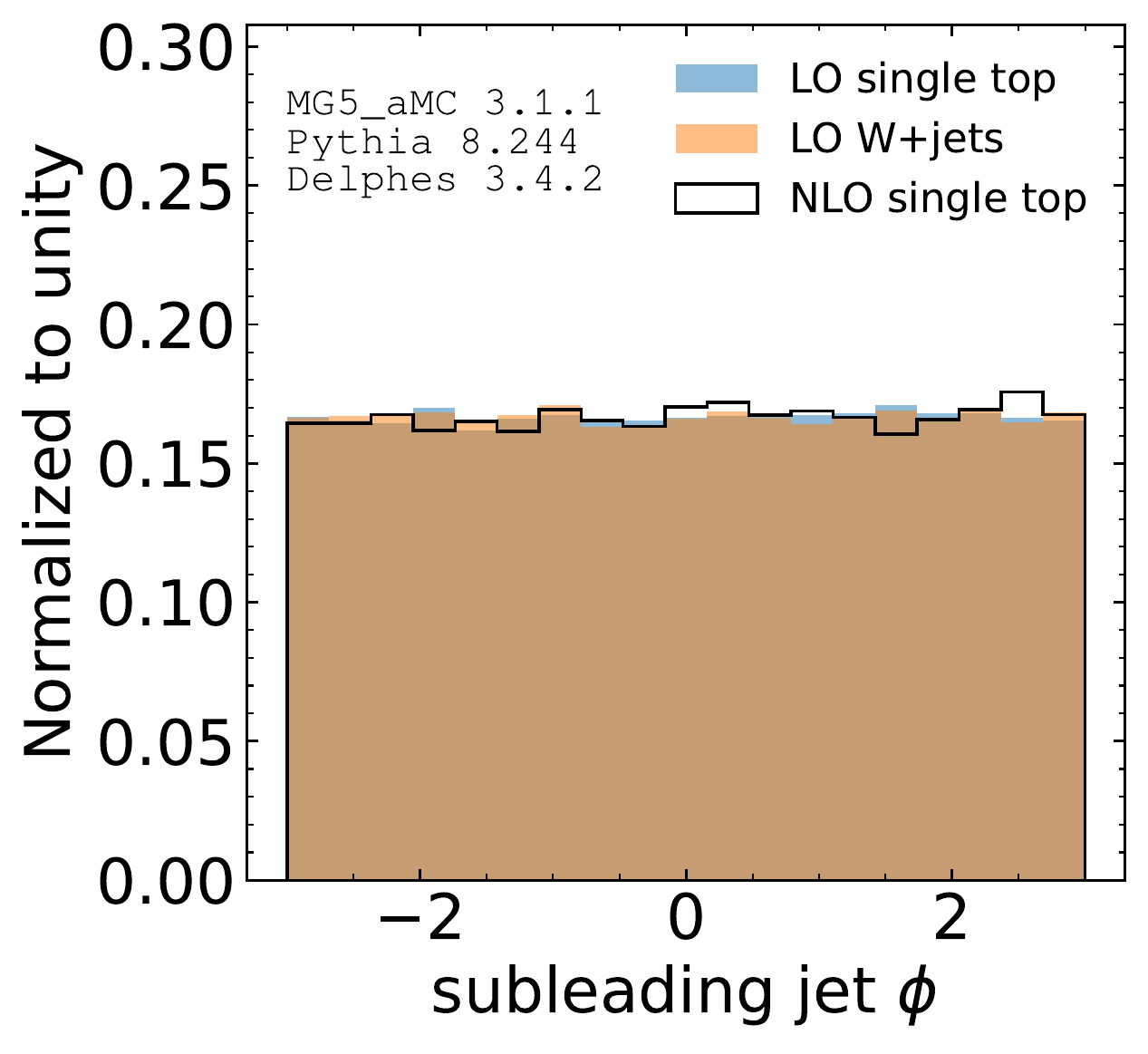}\includegraphics[height=0.25\textwidth]{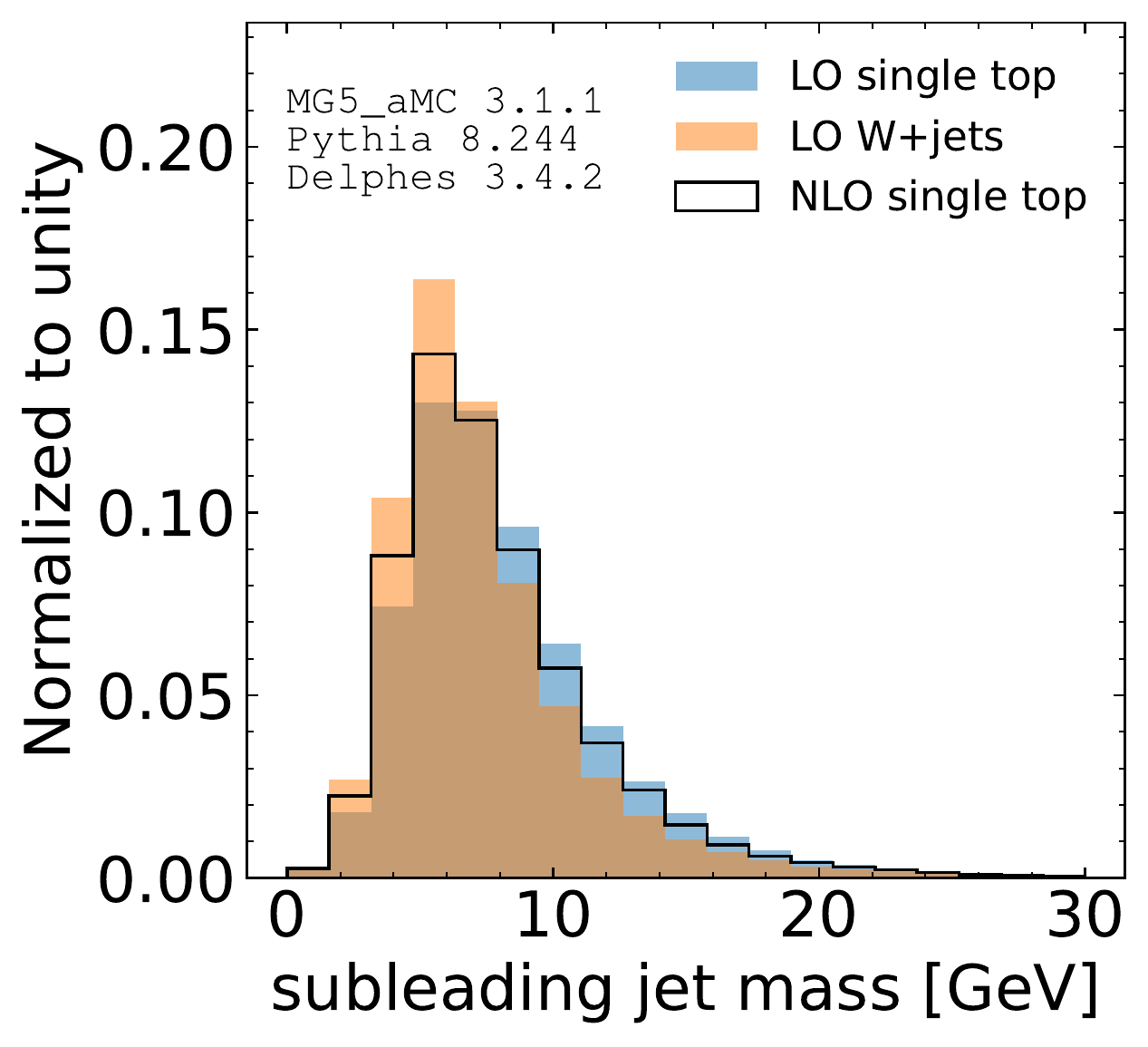}\includegraphics[height=0.25\textwidth]{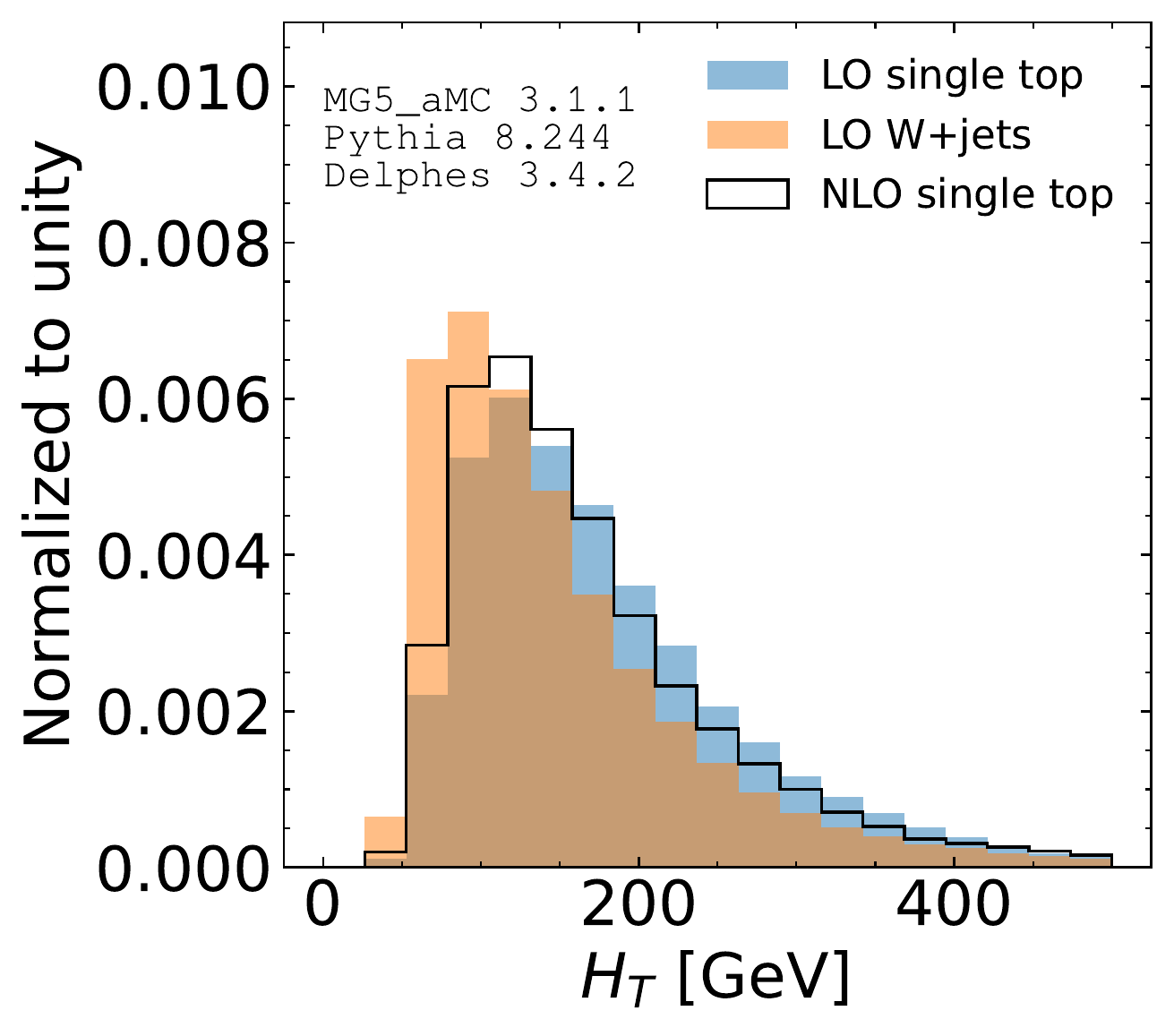}
\caption{The 12 inputs used to train a classifier to distinguish single top events from $W$+jets events.}
\label{fig:stinputs}
\end{figure}

For the machine learning, events are represented by 12 numbers: the three-momentum of the muon, the four-momentum of the leading two jets, and the scalar sum of the transverse momenta of all jets ($H_T$).  Momenta are specified by $p_T$, $\eta$, and $\phi$.  Histograms for each of the observables for single top $t$-channel and $W$+jets are shown in Fig.~\ref{fig:stinputs}.  The jet $p_T$ spectra are harder for single top compared with $W$ jets and the muons (jets) tend to be more central (forward) for single top compared with $W$+jets.

\begin{figure}[h!]
\centering
\includegraphics[height=0.25\textwidth]{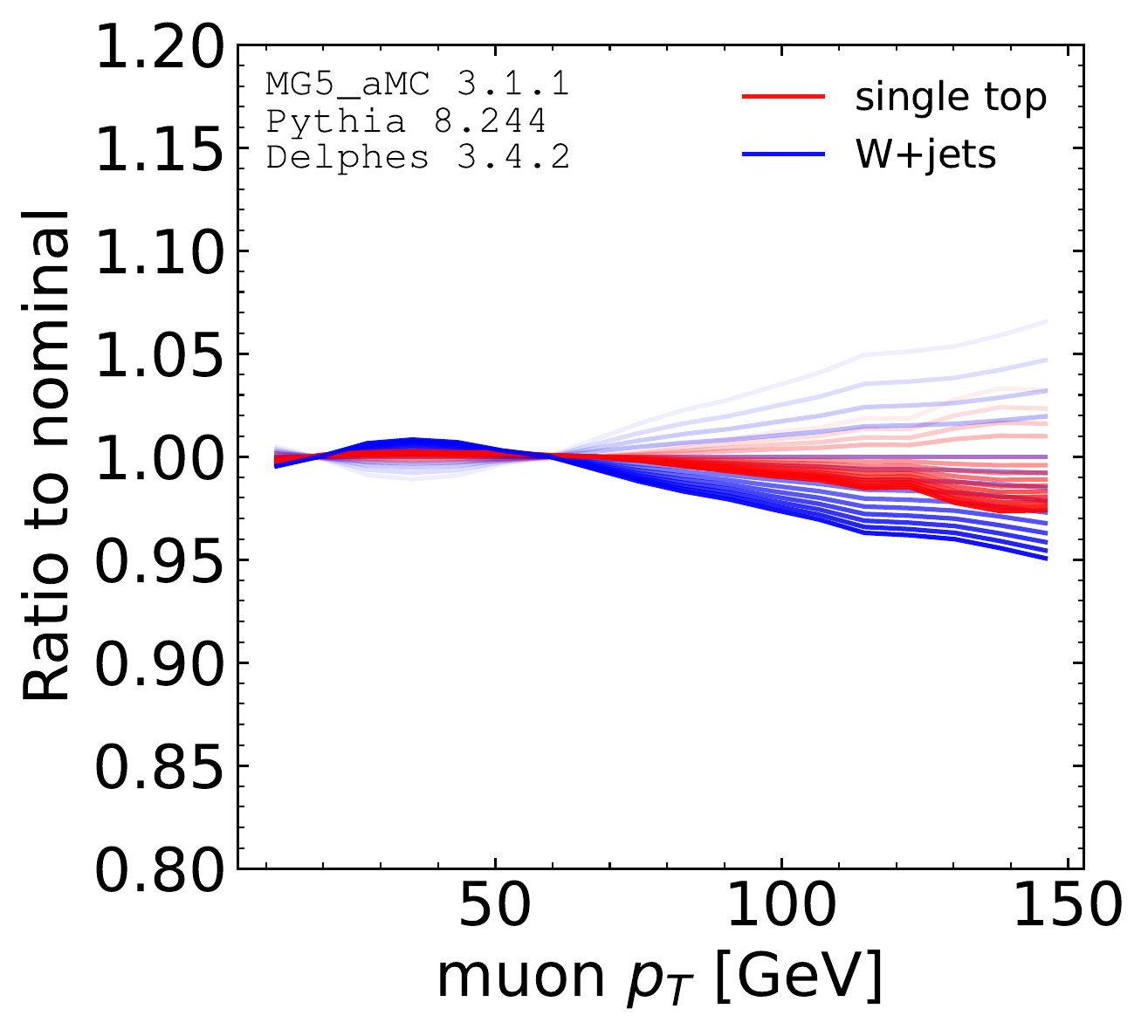}\includegraphics[height=0.25\textwidth]{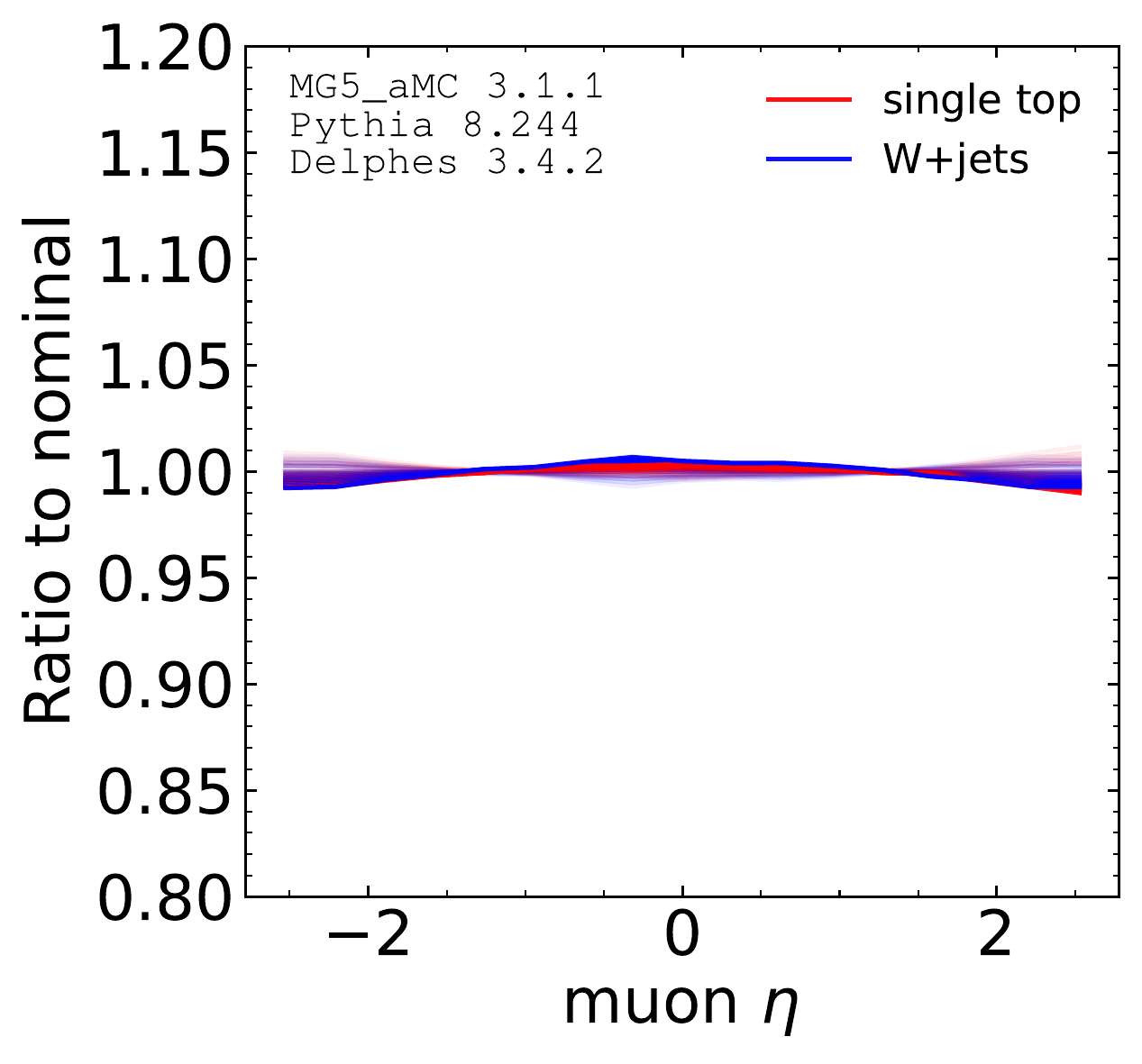}\includegraphics[height=0.25\textwidth]{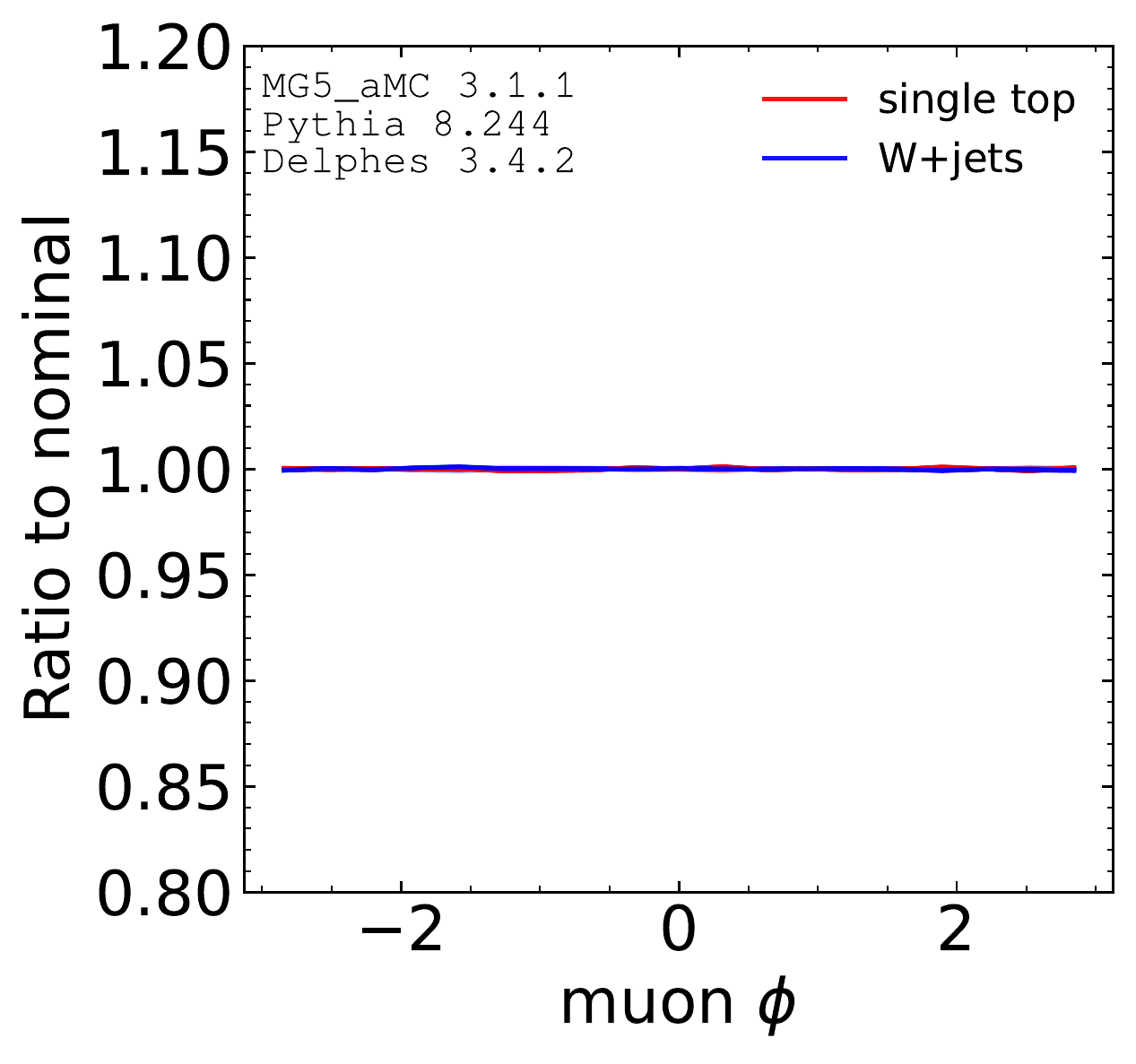}\\
\includegraphics[height=0.25\textwidth]{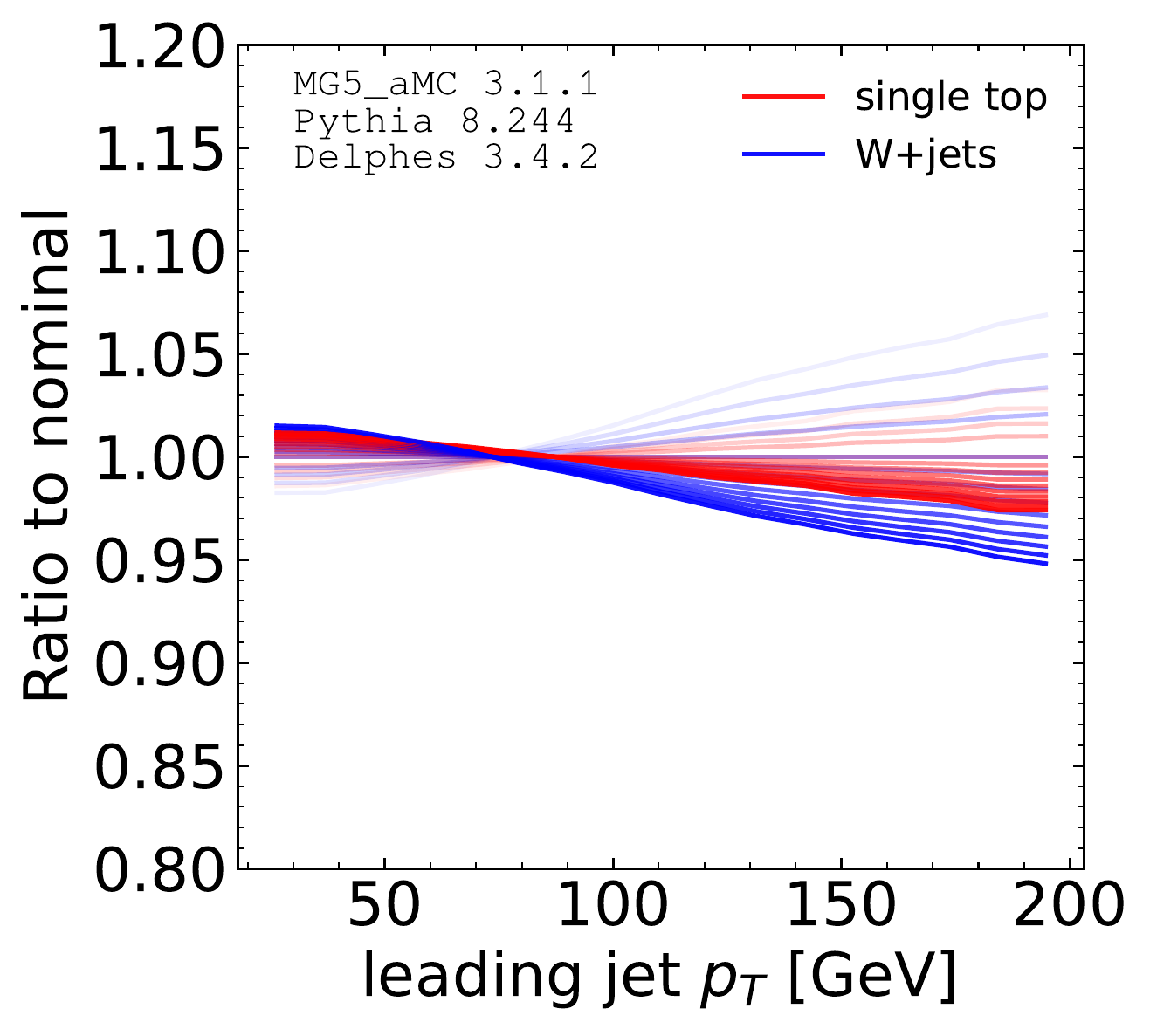}\includegraphics[height=0.25\textwidth]{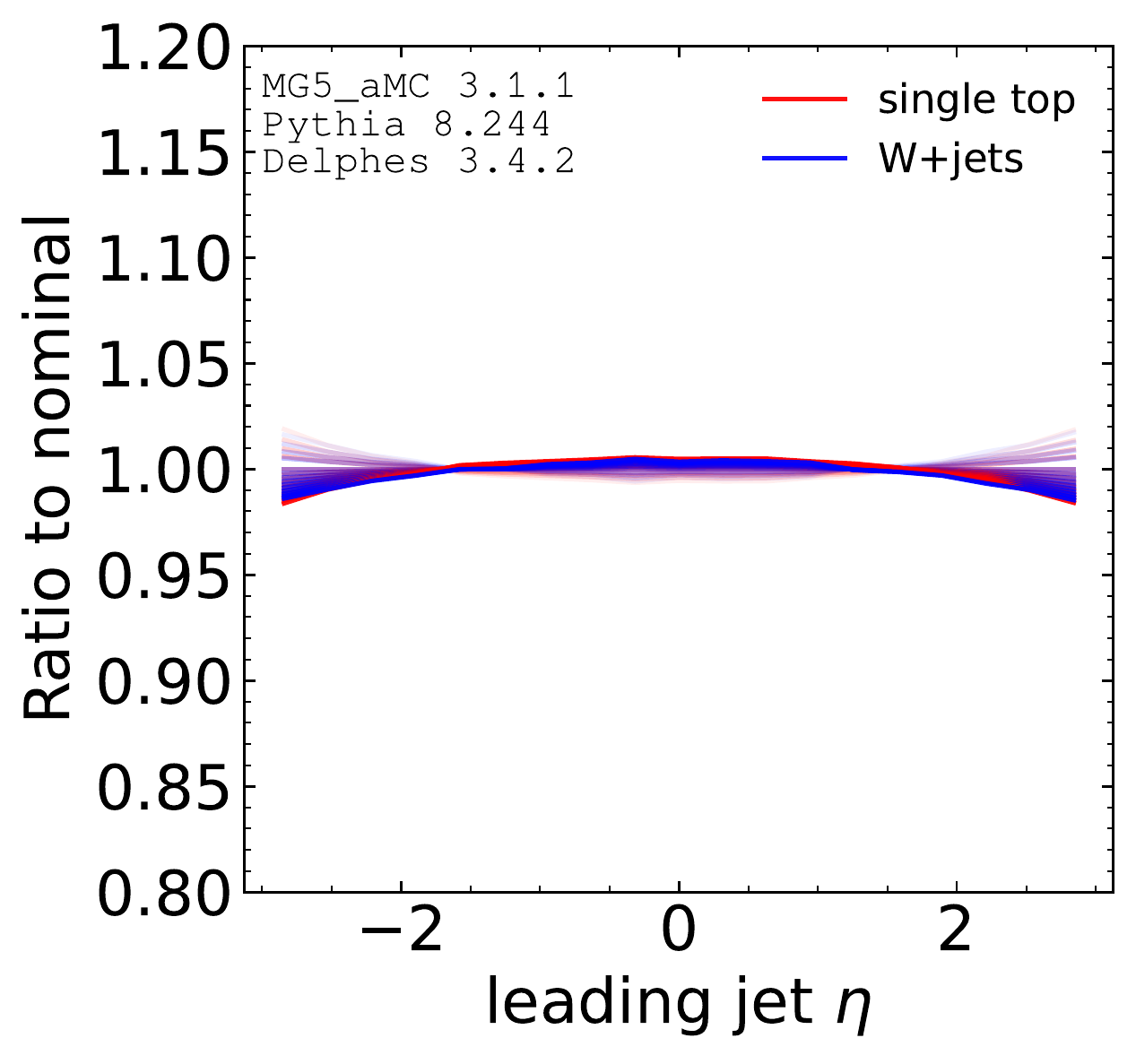}\includegraphics[height=0.25\textwidth]{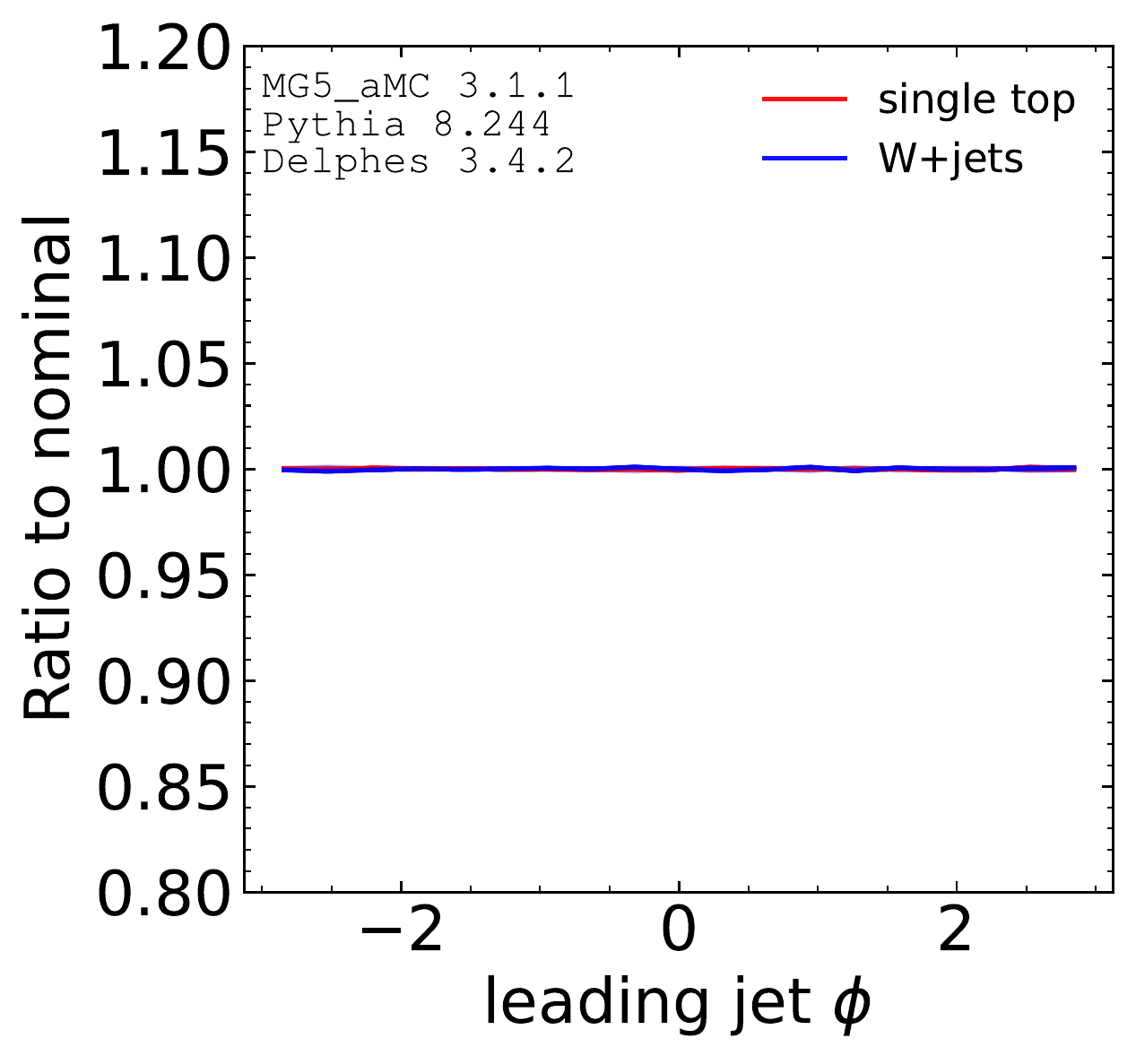}\\
\includegraphics[height=0.25\textwidth]{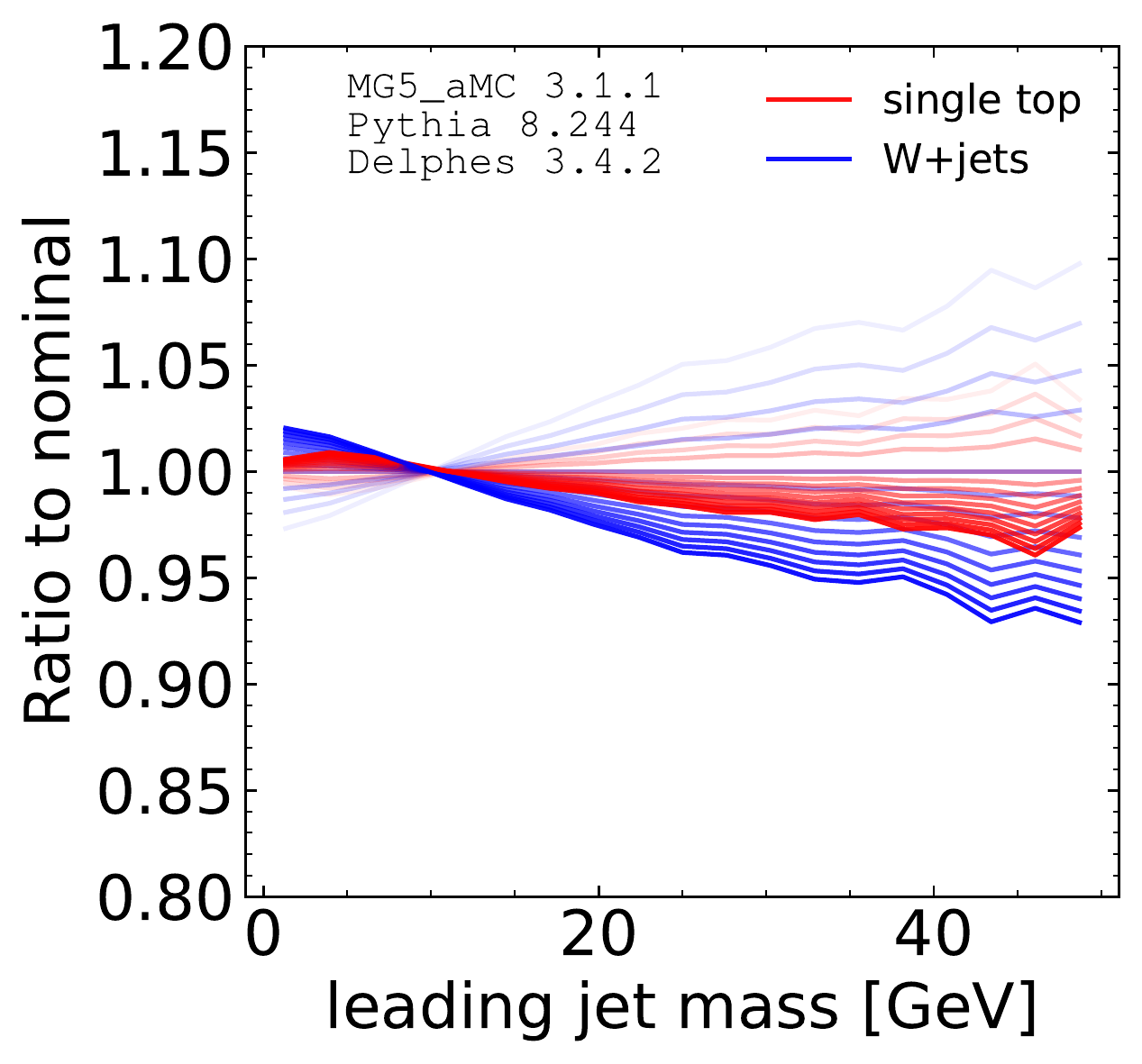}\includegraphics[height=0.25\textwidth]{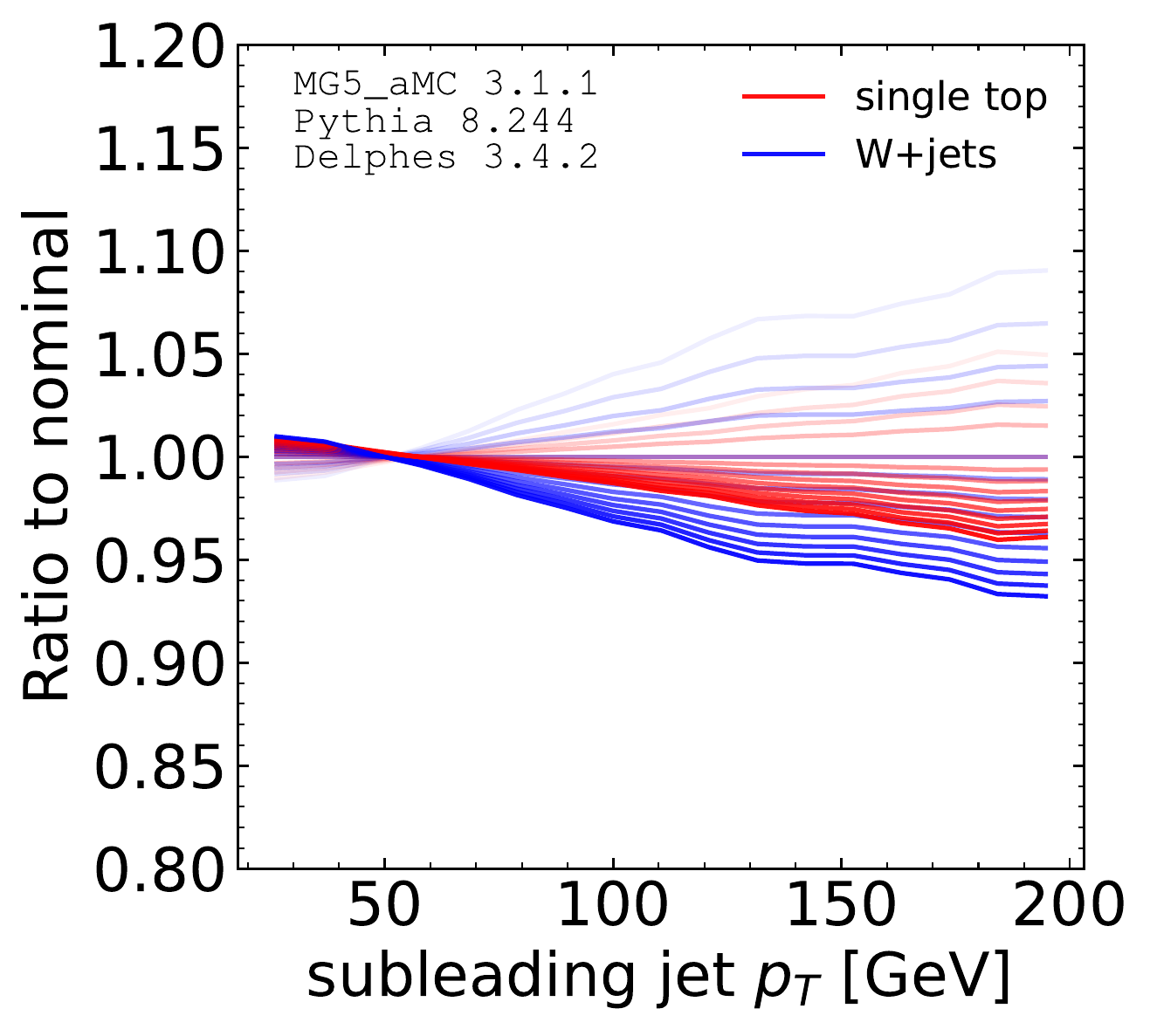}\includegraphics[height=0.25\textwidth]{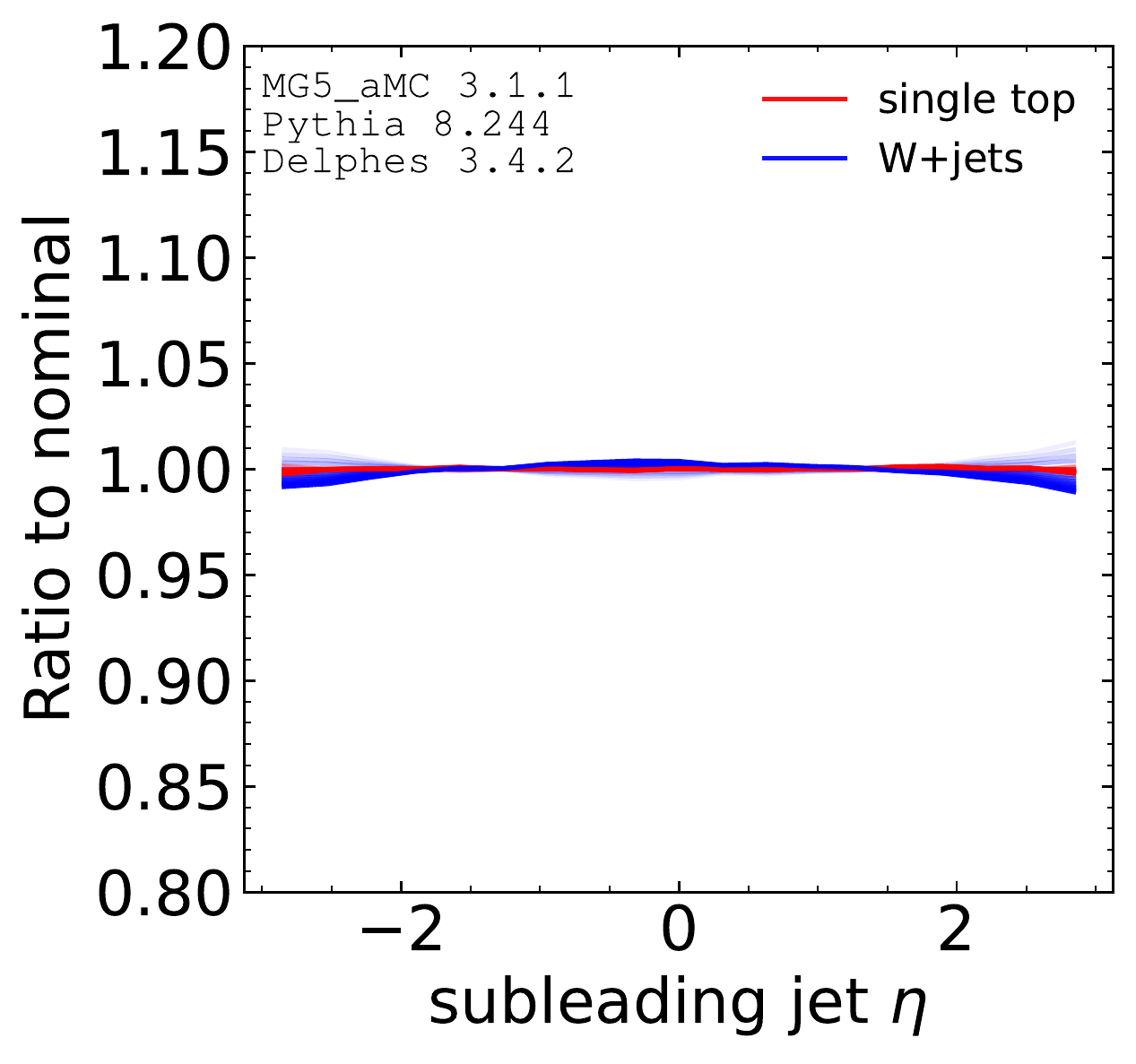}
\includegraphics[height=0.25\textwidth]{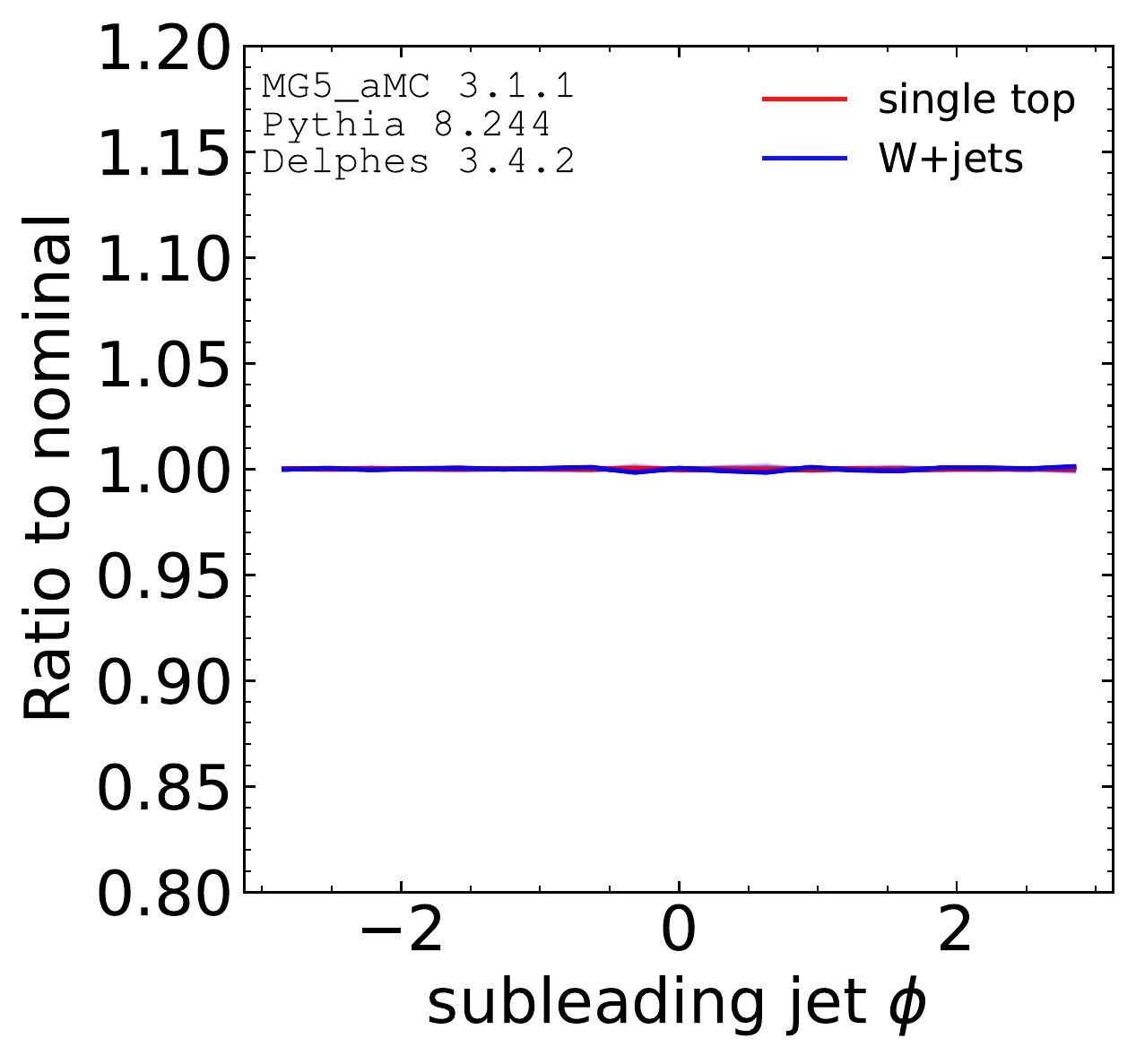}\includegraphics[height=0.25\textwidth]{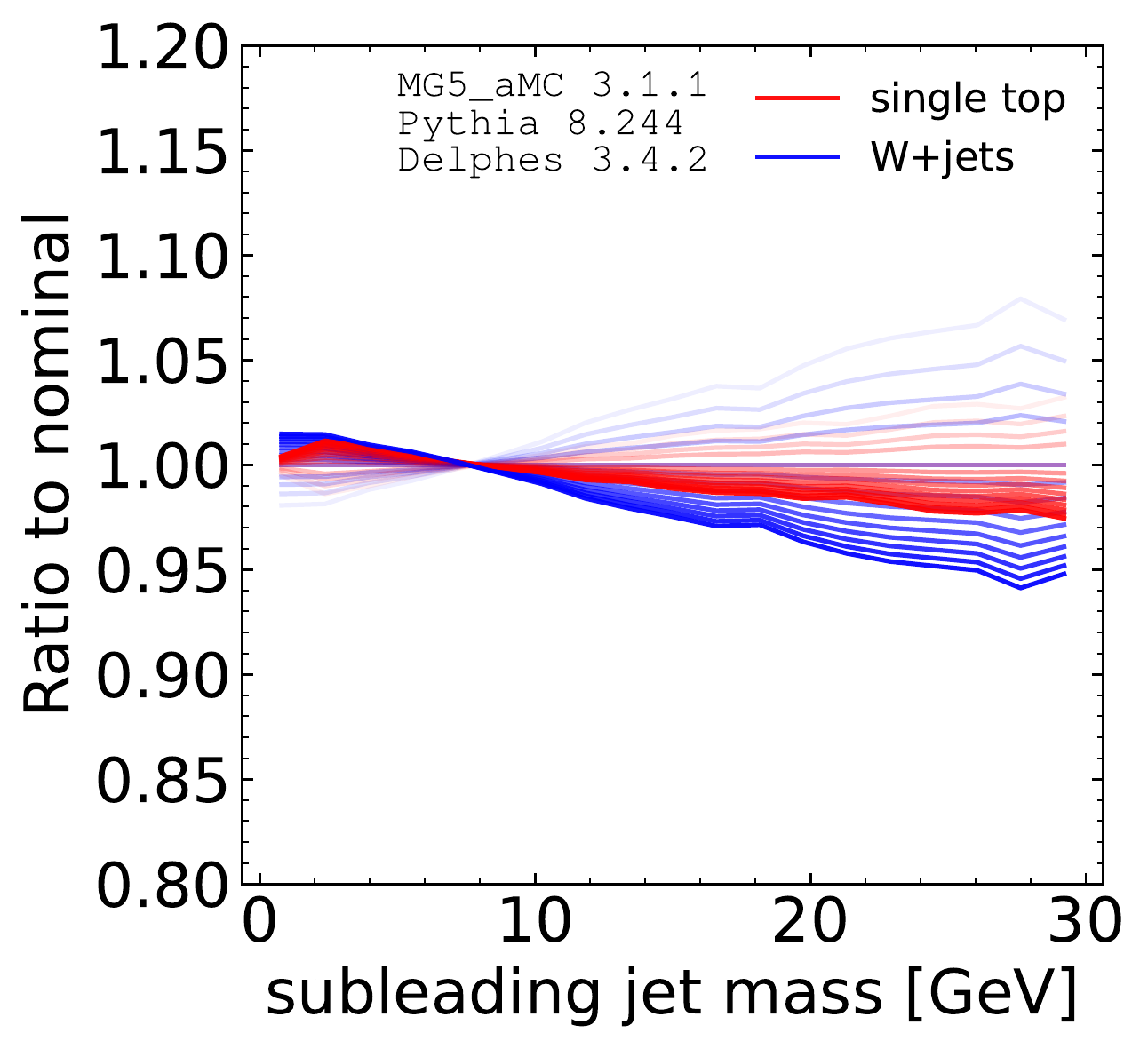}\includegraphics[height=0.25\textwidth]{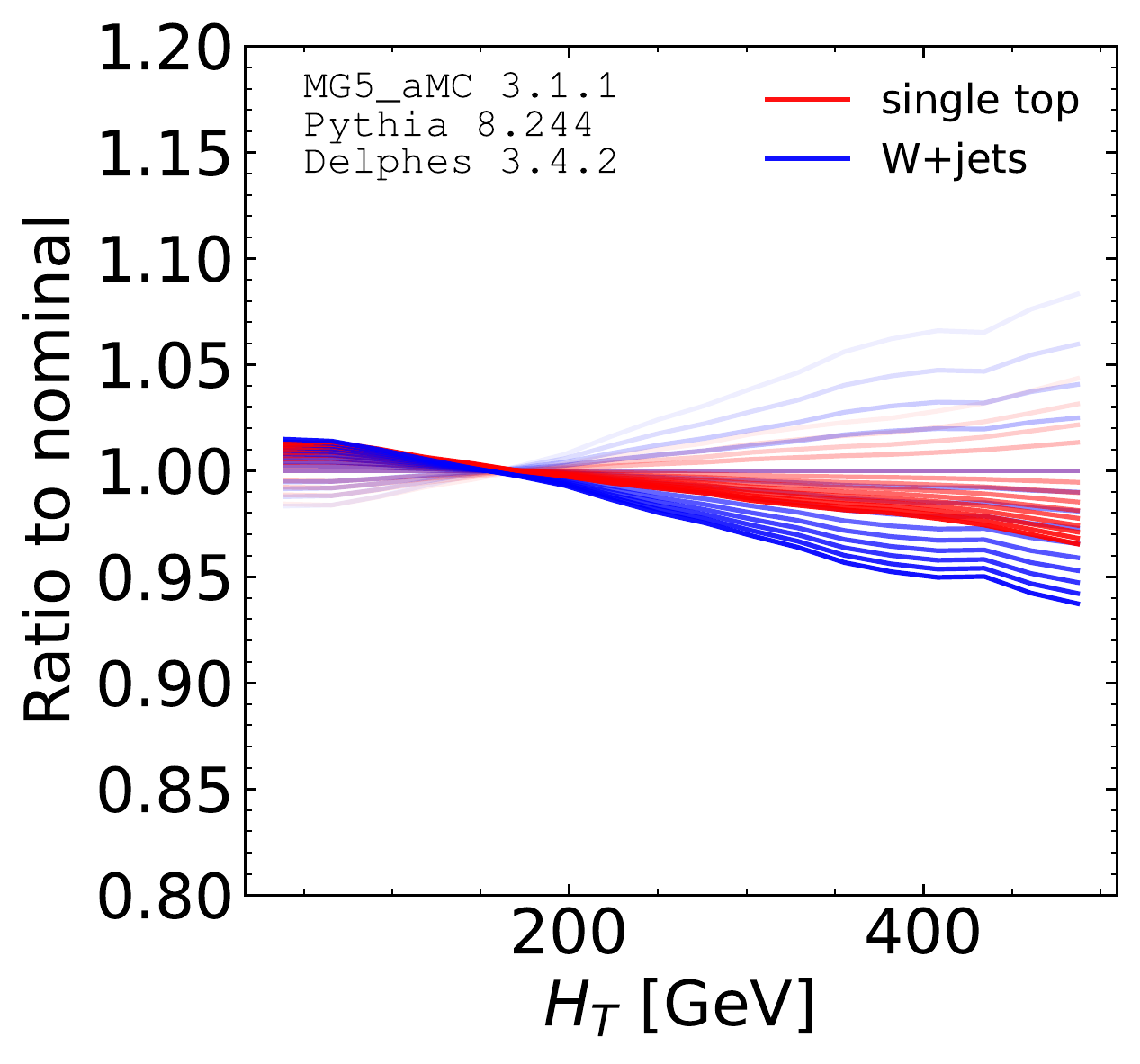}
\caption{The impact of factorization scale variations by a factor of $1/2$ and $2$, in increments of 0.1 (lighter colors are lower scales).}
\label{fig:stinputs2}
\end{figure}

The impact of factorization scale variations is shown in Fig.~\ref{fig:stinputs2}.  All variations are normalized to unity, as the impact on the total cross section is not relevant for per-event classification performance.  As expected, the variation for all $\phi$ observables is negligible and the biggest variation occurs for the transverse momenta.

\begin{figure}[h!]
\centering
\includegraphics[height=0.65\textwidth]{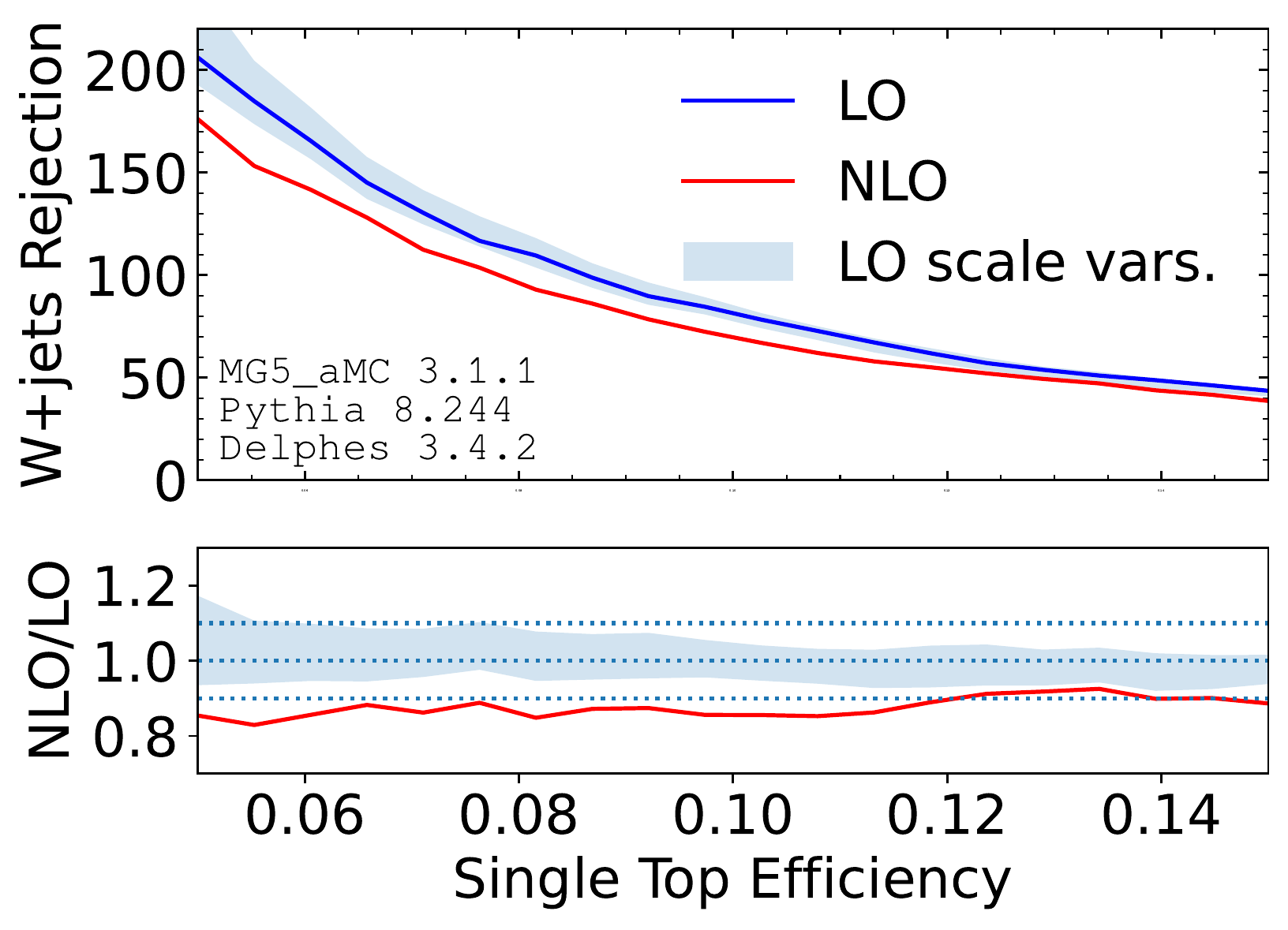}\\\includegraphics[height=0.65\textwidth]{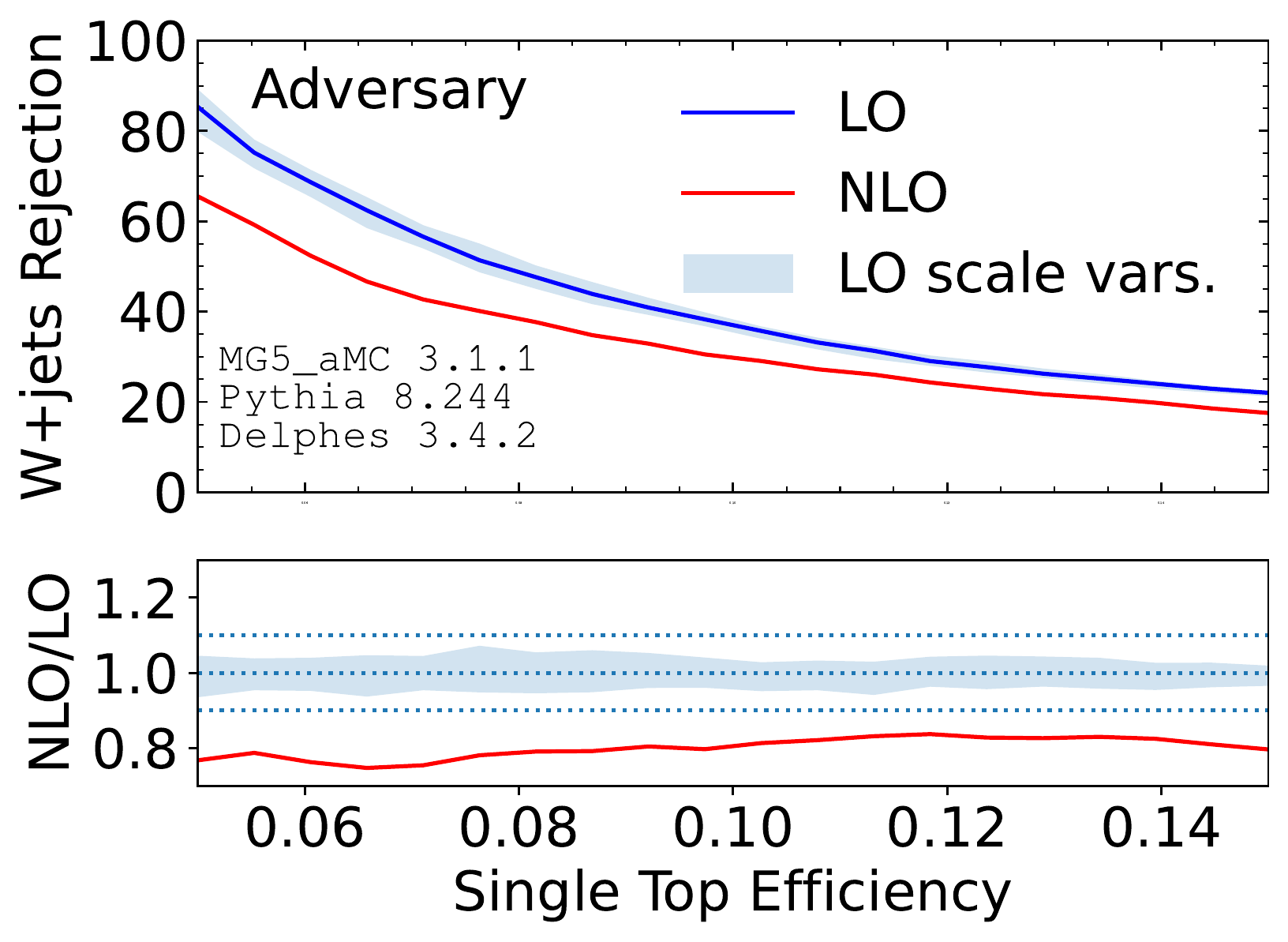}
\caption{Top: the performance of the nominal $t$-channel single top versus $W$+jets classifier. The blue band represents the uncertainty estimated by varying the factorization scale by $\frac{1}{2}$ and 2 at LO.  Bottom: the same as the top, but for the adversarially trained classifier. Adversarial training only reduces the difference in performance to factorization scale variations, not the difference to NLO, indicating that adversarial training provides a reduced \emph{estimate} of the true uncertainty, which does not translate to a reduction in the true uncertainty.}
\label{fig:rocsst}
\end{figure}

The default performance for a classifier trained to distinguish single top events from $W$+jets events is shown in the top plot of Fig.~\ref{fig:rocsst}.  The $W$+rejection at a single top efficiency of 10\% is about 75, with about 15\% lower rejection when the single top is simulated at NLO.  Similarly to the fragmentation modeling, an adversarial network is also trained to reduce the sensitivity to factorization scale variations.  Since the scale variation is now continuous, the adversary is trained using the mean squared error:

\begin{align}
\label{eq:advloss2}\nonumber
    L[f,g]=&-\sum_\mu\Bigg[\left(\sum_{i\in \text{LO $t$-chan}} w_i(\mu)\,\log(f(x_i))-\sum_{i\in\text{LO $W$+jets}}w_i(\mu)\,\log(1-f(x_i))\right)\\
    &\hspace{15mm}+ \lambda \,\sum_{i\in \text{LO $t$-chan}} w_i\,(g(f(x_i),y_i)-\mu)^2\Bigg]\,,
\end{align}
where $\mu$ is the relative factorization scale.  For each event, we can vary the factorization scale through per-event weights $w_i$ and we use values $\mu\in\{0.5,0.6...,1.9,2\}$ for each event.  All hyperparameters are the same as for the fragmentation modeling example shown in the previous section.  The performance of the adversarially trained classifier is shown in the bottom plot of Fig.~\ref{fig:rocsst}.  The overall performance is reduced by about a factor of 2 and the sensitivity to factorization scale variations is also significantly reduced by a factor of two or more. While the narrower uncertainty bands may give the impression that the uncertainty has been reduced, in truth the difference between the LO and NLO curves is about the same or bigger than in the nominal case.  This means that the `true' uncertainty would be significantly underestimated using the adversarially trained approach.

\clearpage

\section{Conclusions and Outlook}
\label{sec:conclusion}

Decorrelation is a powerful tool for ensuring that machine learning classifiers can be used in practice to enhance analysis sensitivity.  However, this tool must be used with caution.  We have shown that decorrelation methods may result in significantly underestimated theory uncertainties when using standard approaches to theory uncertainty estimation.  In the cases we explored, the estimated uncertainty uses two samples while the `true' uncertainty relies on a third sample that is not part of the training.  One could potentially incorporate the third sample into the decorrelation procedure, but there will always be another variation that is not part of the training as long as the full theory uncertainty decomposition is not known.  Until we know the complete set of theory nuisance parameters, it seems prudent to not decorrelate away these uncertainties.  

While this paper explicitly studied the case for decorrelation, this cautionary tale remains relevant for other uncertainty or inference aware machine learning approaches~\cite{Ghosh:2021roe,Wunsch:2020iuh,Elwood:2020pik,Xia:2018kgd,deCastro:2018mgh,Charnock_2018,Alsing:2019dvb,lukas_heinrich_2020_3697981,Brehmer:2019xox,Brehmer:2018hga,Brehmer:2018kdj,Brehmer:2018eca,Nachman:2019dol} if they are being considered for such theory uncertainties.

\section*{\label{sec::acknowledgments}Acknowledgments}

We are grateful to Yi-Lun Chung for producing the fragmentation variation samples used in Sec.~\ref{sec:twopoint}.  We thank Kingman Cheung, Shih-Chieh Hsu, Tilman Plehn, David Rousseau, David Shih, Michael Spannowsky, and Daniel Whiteson for useful discussions and comments on the manuscript. BN was supported by the Department of Energy, Office of Science under contract number DE-AC02-05CH11231.  AG was supported by the U.S. Department of Energy (DOE), Office of Science under Grant No. DE-SC0009920.

\clearpage
\appendix

\section{Training with $\lambda=0$}
\label{sec:lamb0}

\begin{figure}[h!]
\centering
\includegraphics[height=0.65\textwidth]{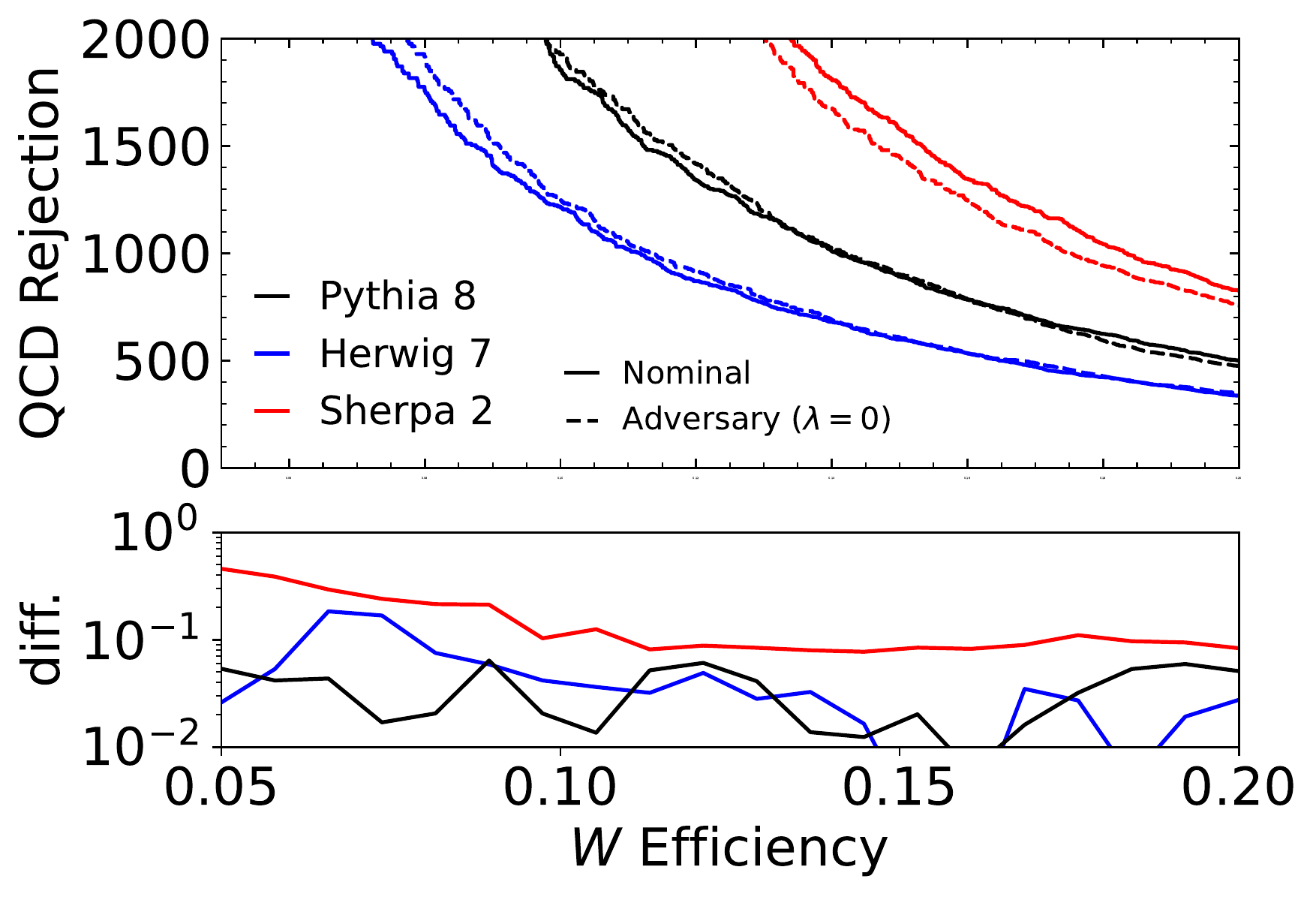}
\caption{ROC curves for the fragmentation modeling example (Sec.~\ref{sec:twopoint}) with the nominal configuration and for an adversary with $\lambda=0$.  The lower panel is the absolute relative difference for each sample between the nominal and adversarial setup.}
\label{fig:lam0_1}
\end{figure}

\begin{figure}[h!]
\centering
\includegraphics[height=0.65\textwidth]{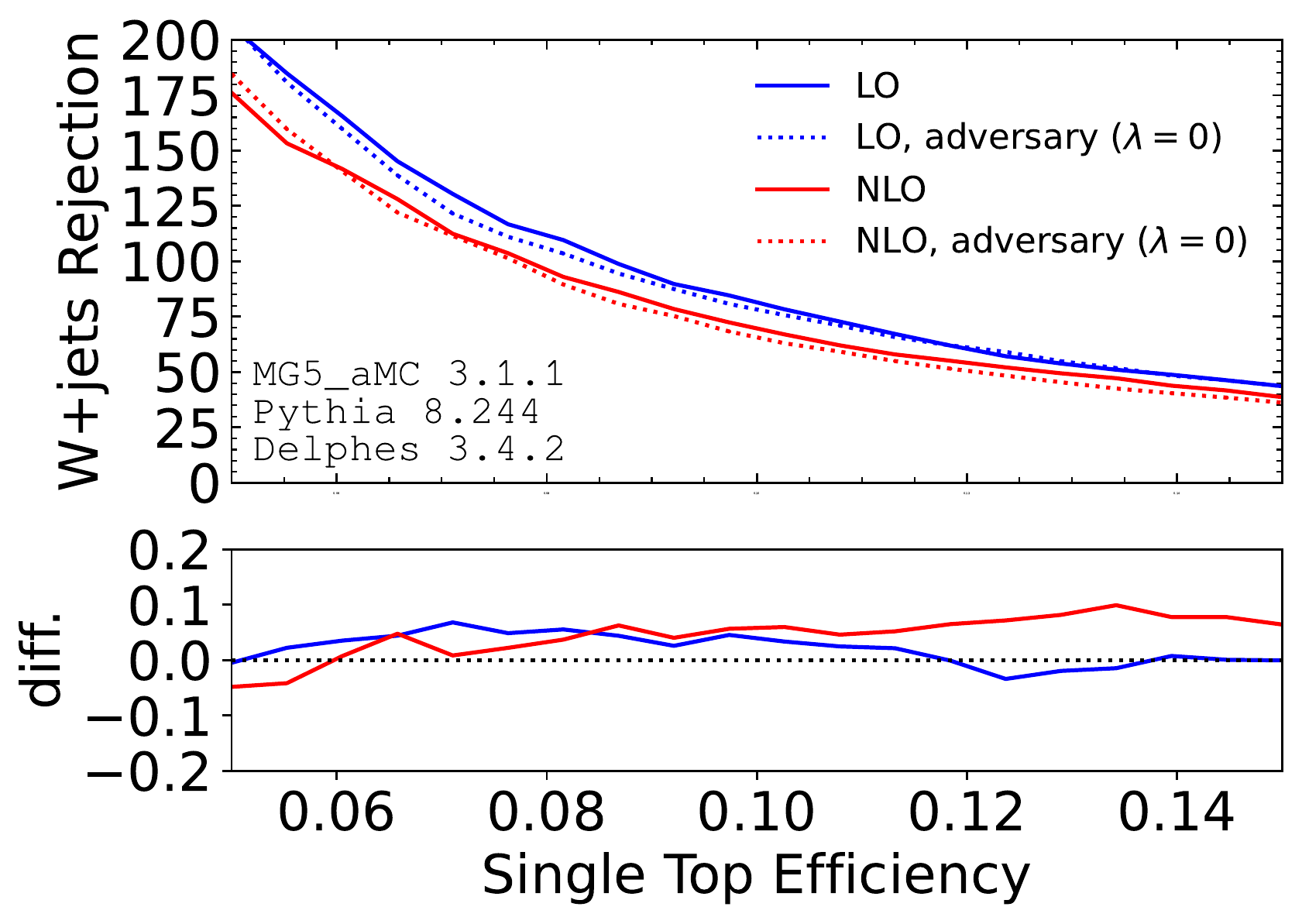}
\caption{ROC curves for the $t$-channel single top example (Sec.~\ref{sec:cont}) with the nominal configuration and for an adversary with $\lambda=0$. The lower panel is the relative difference for each sample between the nominal and adversarial setup.}
\label{fig:lam0_2}
\end{figure}

Figures~\ref{fig:lam0_1} and~\ref{fig:lam0_2} show the impact of using the adversarial setup, but with $\lambda=0$, i.e. the adversary is turned off.  The only difference with respect to the nominal configuration is that \textsc{Pythia} and \textsc{Herwig} (factorization scale variations) are used instead of just \textsc{Pythia} ($\mu=1$) for the nominal for the two-point (continuous) uncertainty example.

\clearpage

\bibliographystyle{JHEP}
\bibliography{main,HEPML}

\end{document}